\documentclass{aa}  
\usepackage{graphicx}
\usepackage{txfonts}
\usepackage{subcaption}

\usepackage[switch]{lineno}
\usepackage{datetime}
\usepackage{setspace}
\usepackage[normalem]{ulem}
\usepackage{soul}
\usepackage{hyperref}

\begin{document}

   \title{A fresh perspective on the 3D dynamics of Tycho's supernova remnant: Ejecta asymmetries in the X-ray band }

   \author{L. Godinaud\inst{1}
          \and
          F. Acero \inst{2,3}
          \and 
          A. Decourchelle \inst{2}
          \and 
          J. Ballet \inst{2}
          }

   \institute{Université Paris Cité, Université Paris-Saclay, CEA, CNRS, AIM, F-91191, Gif-sur-Yvette, France \\
              \email{leila.godinaud@cea.fr}
              \and
            Université Paris-Saclay, Université Paris Cité, CEA, CNRS, AIM, 91191, Gif-sur-Yvette, France \\
              \email{fabio.acero@cea.fr}
              \and
            FSLAC IRL 2009, CNRS/IAC, La Laguna, Tenerife, Spain \\              
             }

   \date{Submitted May 19, 2023; accepted August 5, 2023}

  \abstract
   {Even 450 years after the explosion of the Type Ia SN 1572, the dynamics of the Tycho supernova remnant (Tycho's SNR) can offer keys to improving our understanding of the explosion mechanism and the interaction of the remnant with the interstellar medium.}
   {To probe the asymmetries and the evolution of the SNR, we tracked the ejecta dynamics using new methods applied to the deep X-ray observations available in the \emph{Chandra} space telescope archive.} 
   {For the line-of-sight (LoS) velocity measurement ($V_{\rm z}$), we used the Doppler effect focused on the bright Si line in the 1.6-2.1 keV band.
   Using the component separation tool called general morphological component analysis (GMCA), we successfully disentangled the red and blueshifted Si ejecta emission. This allowed us to reconstruct a map of the peak energy of the silicon line with a total coverage of the SNR at a 2'' resolution. We then obtained a proxy of the integrated velocity along the LoS.
    For the proper motions in the plane of the sky ($V_{\rm xy}$), we developed a new method, called Poisson optical flow (POF), to measure the displacement of two-dimensional (2D) features between the observations of 2003 and 2009. The result is a field of around 1700 velocity vectors covering the entire SNR.}
   {These exhaustive three-dimensional (3D) velocity measurements reveal the complex dynamics of Tycho's SNR. 
   Our study sheds light on a patchy $V_{\rm z}$ map, where most regions are dominated by the foreground or the background part of the shell. On a large scale, an asymmetry is seen, with the north being dominantly blueshifted and the south redshifted. 
   The proper-motion vector field, $V_{\rm xy}$, highlights different dynamics between the eastern and the western parts of the SNR. The eastern velocity field is more disturbed by external inhomogeneities and the south-east ejecta knot. 
   In particular, a slow-down is observed in the north-east, which could be due to the interaction with higher densities, as seen in other wavelengths.
   The vector field is also used to backtrace the center of the explosion, which is then compared with potential stellar progenitors in the area. The latest \emph{Gaia} DR3 parallax measurements exclude most stellar candidates based on their distances, leaving only stars B and E as possible candidates, at respective distances of 2.53$^{+0.23}_{-0.20}$ kpc and 3.52$^{+2.0}_{-1.0}$ kpc, which are consistent with the expected distance range of the SNR at 2.5-4 kpc.}
   {}


   \maketitle
%

\section{Introduction}

The 450th anniversary of Tycho’s Nova Stella (SN 1572) offers an occasion to revisit the deep X-ray archival observations of the \emph{Chandra} telescope with recent advanced analysis techniques. A type Ia supernova explosion is at the origin of this "new star" observed in November 1572 by Tycho Brahe. Earlier observations had also been recorded by Korean and Chinese astronomers \citep{GreenStephenson2003}. The event is thought to have been a "normal" type Ia based on analysis of the X-ray emitting ejecta \citep{Badenes2006}, further confirmed by the spectroscopy of the observed light echoes of the explosion \citep{Krause2008}. However, the understanding of type Ia supernovae is still subject to debate. Two scenarios are possible: the single-degenerate model (a white dwarf accreting matter from a non-degenerate companion) and the double degenerate model (the explosion coming from the interaction of two white dwarfs).
Centuries after the explosion, these explosion scenarios will influence the type Ia supernova remnant (SNR) and its dynamics \citep{Ferrand2019}. These details can be probed by the ejecta X-ray emission in young ejecta-dominated SNRs.

Contrary to the core collapse SNRs, remnants of thermonuclear supernovae show a  more spherical expansion \citep{Lopez2011}, as observed in Tycho's SNR.
However, some asymmetries can be highlighted by studying the dynamics in detail. Their origin can be innate or acquired: either due to an initial anisotropy in the supernova or related to interactions between the expansion and inhomogeneities in the ambient interstellar medium.
Simulations show that an initial asymmetric explosion will leave an imprint in the SNR hundreds of years later \citep{Ferrand2019, Ferrand2022}. Some high-velocity components seen in the echo light of this SNR could be explained by an aspherical supernova \citep{Krause2008}. The origin of the fast iron and silicon knot in the south-east (SE) is also interpreted as ejecta bullets formed during the explosion \citep{Yamaguchi2017}. In addition, \cite{Sato2019}  showed that clumpiness in the early remnant best explains the current morphology of Tycho's SNR.
However, the environment of Tycho's SNR is known to be inhomogeneous. \cite{Williams2013} found a density gradient based on radio observation, while \cite{Zhou2016} observed in addition a potential molecular cloud in the northwest, also highlighted by \cite{Arias2019}.

To probe these possibilities, studies have been carried out in the X-ray band to follow the SNR's evolution across multiple epochs. The velocity of the forward shock was first studied by measuring the shifts of synchrotron filaments \citep{Katsuda2010, Williams2016, Tanaka2021} following the method of \cite{Katsuda2008}. 
This protocol was then applied to the ejecta \citep{Williams2017, Millard2022} to measure the projected velocity in the plane of the sky. The first direct measurement of the projected velocity in the line of sight (LoS) was realized by \cite{Sato2017a}, using the Doppler effect.
Then \cite{Williams2017} and \cite{Millard2022} combined these two methods to obtain three-dimensional (3D) velocity vectors of around 80 ejecta blobs (combining the two studies).
Based on these dynamics measurements, an east-west asymmetry is observed in the forward shock velocities \citep{Williams2016}, which can be explained by a density gradient \citep{Williams2013}. However, no such asymmetry is seen for the ejecta dynamics in the plane of the sky, except for the fast moving knot in the south-east \citep{Yamaguchi2017}. In the LoS, \cite{Millard2022} used the high resolution grating spectrometer on fifty bright ejecta blobs to highlight a north-south asymmetry, where the northern ejecta is more blue-shifted than the southern regions.

In the case of gratings, only bright blobs can be studied and the number of zones is limited, so there is not enough data to do a statistical study or to consider a 3D reconstruction (x, y, z) of the SNR's expansion.
In previous studies, the 3D nature (x, y, energy) of the X-ray data has not been used to its full potential as, in most cases, the spectral and spatial information are used separately.
New analysis methods can be developed to exploit this wealth of information. For example, a principal component analysis was used by \cite{Warren2005} to find interesting regions to study. \cite{Iwasaki2019} used unsupervised deep learning to propose a more sophisticated decomposition of the supernova remnant Tycho. 
In this article, we use the tool General Morphological Component Analysis \citep{Bobin2015, Picquenot2019}. The general idea is to do a blind source separation on an X-ray data cube and retrieve the components with their common spectral signatures and provide  the spectrum and associated image of each component as the output. It has been used to study the Cassiopeia A SNR in \cite{Picquenot2021} to highlight some redshift and blueshift asymmetries of individual emission lines, as well as in SNR N103B to reveal a double-ring structure in the ejecta component \citep{Yamaguchi2021}.

The objectives of the current paper are to provide a velocity vector field of the ejecta to study the 3D dynamics of the entire Tycho's SNR.
With this aim, we propose novel methods to study the 3D ejecta expansion. We separately analyze  the velocity in the LoS, $V_{\rm z}$, and the velocity in the plane of the sky, $V_{\rm xy}$. 
First, we present the data from the \emph{Chandra} telescope and the new tools we applied in Sections \ref{Section:Observations and data reduction} and \ref{Section:Data analysis methods} .
We obtain a complete map of the peak energy for the silicon line and of the redshift in the LoS (see Section \ref{Section:Line of sight velocities}), along with around 1700 proper motions in the plane of the sky (see Section \ref{Section:Plane of the sky velocities}). This gives precise information on the dynamics asymmetries, an evaluation of the center of the explosion to search for a potential progenitor, and clues toward 3D reconstruction, as discussed in Section \ref{Section:Discussion}.

In this paper, we assume that the distance of Tycho's SNR is 3.5 kpc. A complete review of the distance is given by \cite{Hayato2010} and we use this value to be consistent with the results of \cite{Williams2017}. 
For the center of the explosion used as a reference to measure a radius in the plane of the sky, we adopted the value we found (see Section \ref{subsection : Center of the explosion from the vector field} ) of R.A. 00$^h$25$^m$20$^s$.79  and Dec. 64°08'09".04 .

We also use the following conventions: the velocity in the plane of the sky is called proper motion (hereafter, $V_{\rm xy}$). In the LoS, the velocity measured with the Doppler effect is named $V_{\rm z}$, which is positive away from us.


\section{Observations and data reduction}
\label{Section:Observations and data reduction}

The Tycho SNR has been observed multiple times by the \emph{Chandra} X-ray telescope, in particular, in 2009 with a deep observation of 734 ks with nine observations in a month. We also use the observation from 2003 with around 145 ks of exposure time. All these observations are summarized in Table \ref{table:1}.

\begin{table}[h!]
\caption{\footnotesize \emph{Chandra} observations used in this study}             
\label{table:1}      
\centering                          
\begin{tabular}{c c c}        
\hline\hline                 
ObsID & Date (YYYY/MM/DD) & Exposure time (ks) \\    
\hline                        
3837  & 2003/04/29 & 145.6  \\      
\hline   
10093  & 2009/04/13 & 118.4  \\
10094  & 2009/04/18 & 90.0  \\
10095  & 2009/04/23 & 173.4  \\
10096  & 2009/04/27 & 105.07  \\
10097  & 2009/04/11 & 107.4  \\
10902  & 2009/04/15 & 39.5  \\
10903  & 2009/04/17 & 23.9  \\
10904  & 2009/04/13 & 34.7  \\
10906  & 2009/05/03 & 41.1  \\
\hline                                   
\end{tabular}
\end{table}

In our analysis, the new methods and their inputs are different for the $V_{\rm xy}$ and $V_{\rm z}$ velocities. We must therefore adapt the binning of our data cube (RA, DEC, E) according to the problem.

First, for the $V_{\rm z}$ velocities, we used the GMCA component separation method. This algorithm requires a data cube as its input and high statistics. We focused on the deep 2009 data set and stacked all the observations of the year. 
This method allows us to study the Doppler effect on the silicon line and deduce the velocity. So, we use the native energy binning of 14.6 eV and a spatial binning of 2". It is four times the native spatial binning, meant to obtain a high number of counts in all voxels.
    
For the proper motion $V_{\rm xy}$,  we measured very small shifts between two images from 2003 and 2009. Here, the data cubes are stacked across energy between 0.5 keV and 7 keV to obtain the images. We used the native spatial binning of \emph{Chandra} ($0\farcs5$) to obtain more details.

\begin{figure*}
\centering
\includegraphics[scale=0.5, trim = 0 0 0 30, clip=true]{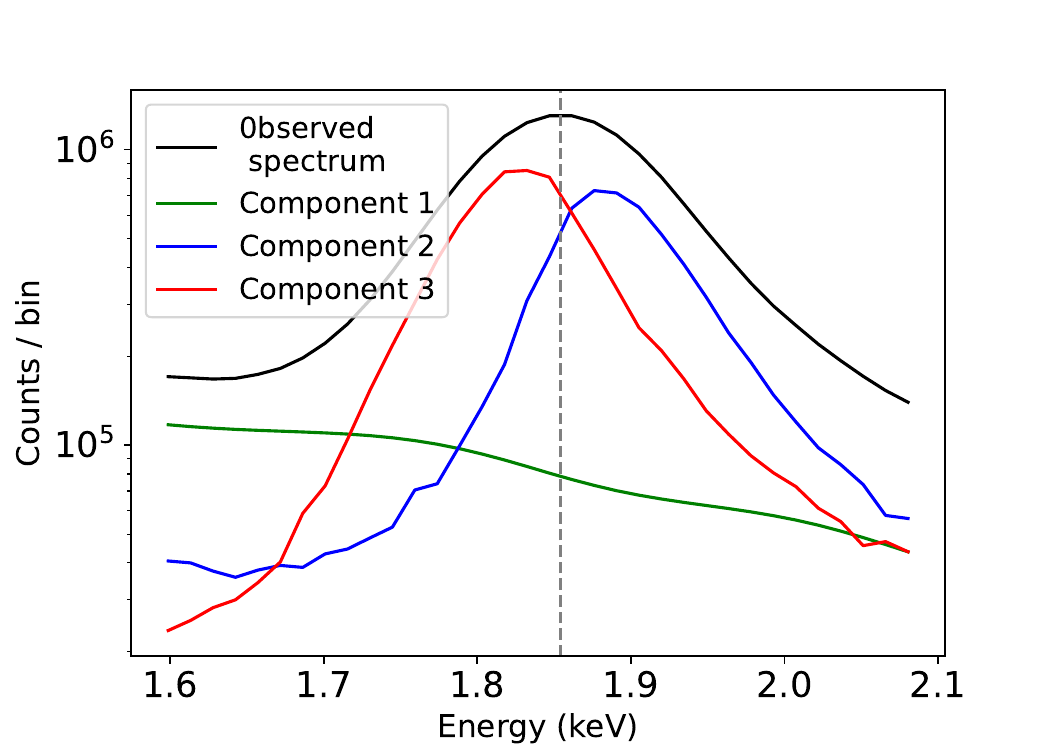}
\includegraphics[scale=0.54, trim = 80 30 0 30, clip=true]{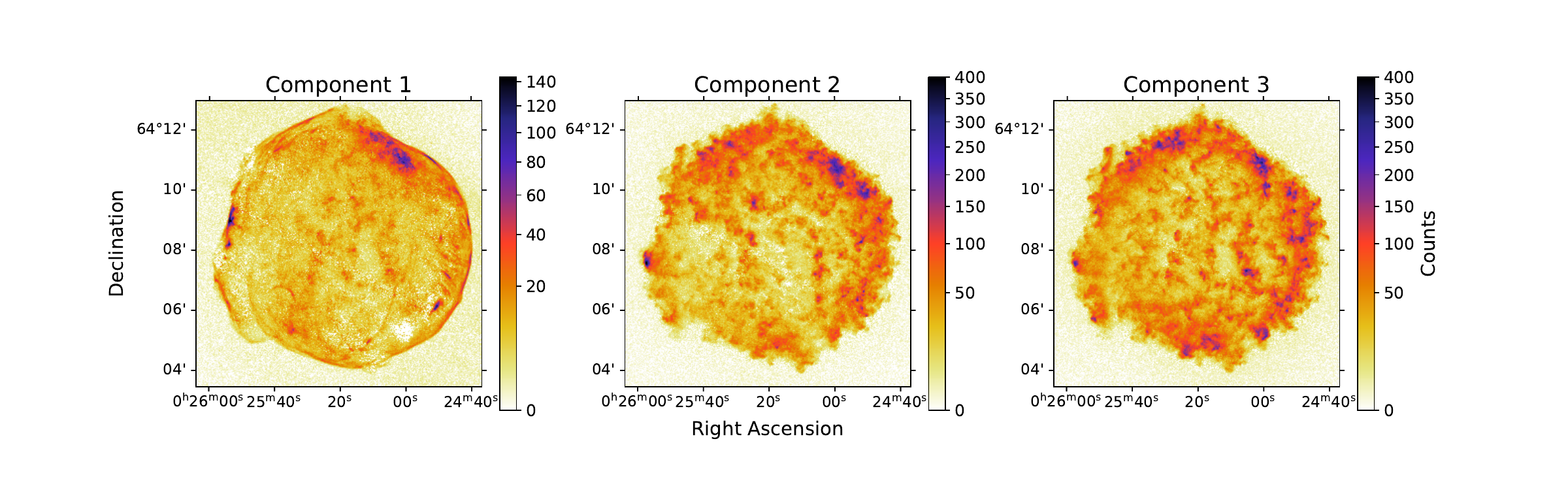}
\caption{\footnotesize GMCA outputs for the data cube of Tycho's SNR in the Si band (1.6 - 2.1 keV). \emph{Top:}  Observed spectrum for the all SNR (in black) compared to the three spectra found by GMCA. \emph{Bottom :} Images associated to the three spectral components found by GMCA. The exposure map was only corrected in the output images, not in the GMCA inputs. Based on the morphology and spectra of the outputs, we interpret the decomposition as follow: the first component corresponds to the continuum, mostly the synchrotron emission. The two others are the thermal emission of the ejecta, with component 2 being redshifted and  component 3 being blueshifted.} 
\label{FigGMCA}
\end{figure*}

Despite the good absolute astrometry of \emph{Chandra}, an image registration of each observation with respect to a reference observation allows for a more accurate astrometry. We note that we could not use the astrometric corrections from \cite{Tanaka2021}, as the data currently available in the archive were reprocessed in late 2020 and are not the same as used in this prior study.
The current reprocessing (\textit{repro5}) comes with an new calibration which provides an improved astrometry\footnote{\url{https://cxc.cfa.harvard.edu/cda/repro5.html\#aspect1}}.
The procedure is to detect point sources with {\tt wavdetect}, compute transformation matrices by crossmatching common sources via {\tt wcs\_match}, and update the event and aspect solution files via  {\tt wcs\_update} (see \footnote{\url{https://cxc.cfa.harvard.edu/ciao/threads/reproject_aspect}} for more details). All observations have been aligned to a reference observation (ObsID 10095, the deepest observation). Depending on the observation, between 4 and 11 common point sources can be used for the alignment. The maximum offset correction is of the order $0\farcs25$ and the average correction of $0\farcs12$. 
We obtain smaller offset corrections compared to \cite{Tanaka2021},  likely due to the improved astrometry provided by \textit{repro5}.


\section{Data analysis methods}
\label{Section:Data analysis methods}

In this section, we present and describe the innovative tools that we use. For the LoS, we used the GMCA method \citep{Bobin2015} to decompose our data cube into the red and blueshifted ejecta components, which is then be used to estimate $V_{\rm z}$. To measure the proper motion, $V_{\rm xy}$, we developed a new tool, called Poisson optical flow (POF), to track the displacement of 2D features across observations.

\subsection{GMCA tool}
\label{subsection:General Morphological Component Analysis tool}

The data coming from a spectro-imaging telescope such as \emph{Chandra} have a 4D nature (x, y, E, t); here, we use the two spatial dimensions and the energy dimension. To  simultaneously exploit the spatial and spectral information and to extract overlapping physical components, we use the GMCA method. This tool decomposes a cube X into a linear combination of spectra, $\rm{A_i}$, and associated images, $\rm{S_i}$, by resolving the inverse problem :

\begin{equation}
    \rm{X  =  \sum^\emph{n}_{\emph{k} = 1} A_\emph{k}\ S_\emph{k} + N}
.\end{equation}

The parameter \emph{N} is the noise that is dealt with by the algorithm and \emph{n} is the number of components chosen by the user. To choose the best number of GMCA components, we tried various values but we were quickly limited by the intrinsic statistics of the data. If too many components are requested, the image output becomes very noisy, with unrealistic discontinuities in the spectra. To find an optimal number of components, we can use the Akaike information criterion \citep[see Appendix B of][]{Picquenot2019}. This parameter corresponds to the negative log-likelihood with a penalty for an increasing number of degrees of freedom. The minimum of this criterion gives the best number of components. We find that three components is the best to decompose our cube centered on the silicon line.

As in a principal component analysis (PCA), we can see the outputs of GMCA spectra as vectors that can be used as a basis to reconstruct the spectrum in all pixels. The weight associated to each vector in a given pixel is the value of this pixel in the associated GMCA image. To disentangle the components, the algorithm optimizes the spatial and spectral differences between the components jointly in the wavelet domain. 
This method is a blind source separation algorithm, with no prior spectral information and, thus, no bias because of a prior. There is nevertheless an option of spectral initialization. The user can constrain the spectra of one or more components and only the normalization of these spectra will be adapted to solve the inverse problem. So, the shape of the spectra must be optimized by the user before. This option can be useful to retrieve a component hidden because of smaller statistics or to clean other components for leakage.

Figure \ref{FigGMCA} shows the GMCA results for the stacked data cube of 2009 in the 1.6-2.1 keV energy band, corresponding to the silicon line. In this analysis, three components were set for the decomposition. The first is initialized to capture the underlying continuum and the second and third are the ejecta that we will study. To initialise the continuum component we use a power-law spectrum with the same parameters as \cite{Williams2017}: a photon index of 2.6 and an absorbing column density of 6 x 10$^{21}$ cm$^{-2}$. Thus, the inputs here are the number of components (three), the data cube to analyze, and the initialization for the power-law spectrum. 

The outputs shown in Fig. \ref{FigGMCA} can be interpreted as physical emissions despite the blind aspect of the separation method. 
The first component corresponds to the fixed power-law component with the goal to capture the underlying synchrotron emission map.
We can see in the image of component 1, that the algorithm successfully retrieves the synchrotron map characterized by filamentary structures despite being buried under the thermal emission from the Si-dominated ejecta. Some leakage from the thermal emission of the ejecta is possible, specifically in the north-west where the thermal emission is particularly bright. Here, the initialization is necessary because the power-law component is too faint to be detected in a pure blind mode in this restricted energy range. To our knowledge, this is Tycho's first synchrotron map in the 1-2 keV band clean of thermal emission.
While this is beyond the scope of this paper, investigating the synchrotron filament structures at different energies could be useful to characterize the magnetic field properties, as done in \citet{Picquenot2023}, and, in particular, for the synchrotron stripes in the west of the SNR.

The second and third components are associated to the ejecta emission: the spectra correspond to the thermal emission (silicon line and underlying Bremsstrahlung continuum) and the associated images show the clumpy aspect due to Rayleigh-Taylor instabilities. GMCA even succeeds at separating blue and redshifted ejecta, as revealed by the shifted spectral lines in the top panel of Fig. \ref{FigGMCA}. 

For the supernova remnants, GMCA can be very efficient because the physical components that we want to separate such as:\ the synchrotron emission, various ejecta elements (intermediate elements or iron emission), and redshifted or blueshifted ejecta, which have very different spectral and spatial signatures. Nevertheless, one limitation of GMCA is the absence of uncertainties for the outputs.

\subsection{Optical flow for measuring proper motions}
\label{subsection:Optical Flow to measure proper motion}

Optical flow methods are a part of the computer vision research domain, which means all the methods linked to detection or velocity measurements. In our case, it consists of measuring the spatial evolution of the ejecta in the plane of the sky. The goal is to detect small shifts of a few pixels between two images within a six-year time interval in our case. For this end, we suppose that there is no significant morphological variation of the small features that we will track between years.
We note that the angular resolution needs to be comparable between the two epochs for all features (ideally with the same telescope pointing ).

We first tested the library OpenCV \footnote{\url{https://opencv.org}}, which is generally used for daily life images and video analysis, as in detecting and measuring the movement of a car. This library has been applied to X-ray observations in \cite{Sato2018} and \cite{Tsuchioka2021}. There are two steps: first detect some good features to track and then measure their displacement between two images. We obtained good results but all the algorithms were completely black box and without a special optimization for astrophysics. In particular, there are no uncertainties in the outputs, no handling of the Poisson noise, or any difference in exposure maps. So, we decided to develop our tool adapted for Poisson statistics of the X-ray data, namely, the POF tool.

The goal is to measure the shift of a small feature across epochs. The deep 2009 flux map (corrected by its exposure map) was used as the model. We can also smooth it to decrease the noise and limit fluctuations in the model if the statistics are limited. The second image is the observation,  which is not modified at any step of the protocol to maintain the Poisson nature of the signal.
The general idea is captured by the following Eq. \ref{EqcstatLandscape}.

\begin{equation}
\rm{
    L(\Delta \emph{x},\Delta \emph{y})= cstat\left( \frac{I_{Mod}(\emph{x}+\Delta \emph{x}, \emph{y}+\Delta \emph{y} )}{Exp_{Mod}(\emph{x}+\Delta \emph{x}, \emph{y}+\Delta \emph{y} )}\ Exp_{Obs} (\emph{x},\emph{y}),\  I_{Obs}(\emph{x},\emph{y}) \right). }
    \label{EqcstatLandscape}
\end{equation}

We created a small vignette around the feature at position $(x,y)$ in the observation image, I$_{\rm Obs}$, and compared it with the equivalent in the model observation, I$_{\rm Mod}$. The model observation moves in X and Y axes with shifts $\Delta x$ and $\Delta y$, in a zone of exploration. We made a cubic interpolation of the model vignette to do sub-pixel steps (five times smaller than the native pixel). Then, at each position, we evaluated the 2D likelihood L$(\Delta x,\Delta y)$ with the \emph{cstat} statistical function \citep{Cash1979}, adapted to the Poisson statistic.
In this way, we created a complete statistical landscape corresponding to all the explored zone around the feature. The minimum of this landscape corresponds to the most likely displacement where the model and observed vignette overlap.

Next, to precisely measure the shift, we carried out a local 2D fit of the statistical landscape (with a 2D polynomial function of degree of 4) only in an area of 2$\times$2 native pixels around the local minimum. It corresponds to the distance between the minimum and the initial position with sub-pixel precision. Finally, we obtain the proper motion, $V_{\rm xy}$, by dividing this shift by the baseline of six years and supposing a distance of Tycho's SNR of 3.5 kpc. We can also derive the ellipse of uncertainties: it is the cut of the 3D landscape for a \emph{cstat} equal to the minimum plus $\Delta cstat$. For uncertainties at 1 sigma, $\Delta cstat$ is equal to 2.3. It is noticeable that Cstat varies a great deal at the native pixel scale, that is much more than 2.3. The interpolations are necessary to obtain the uncertainties. We present in Appendix \ref{Appendix : POF} some examples of features, statistical landscapes, and profiles from our method. A similar idea was used by \cite{Sato2017b} to measure proper motions in Kepler's SNR. We add the exposure map correction and the interpolation of the statistical landscape to have precise uncertainties.

\begin{figure*}
\centering
\includegraphics[scale=0.4, trim = 80 70 0 100, clip=true]{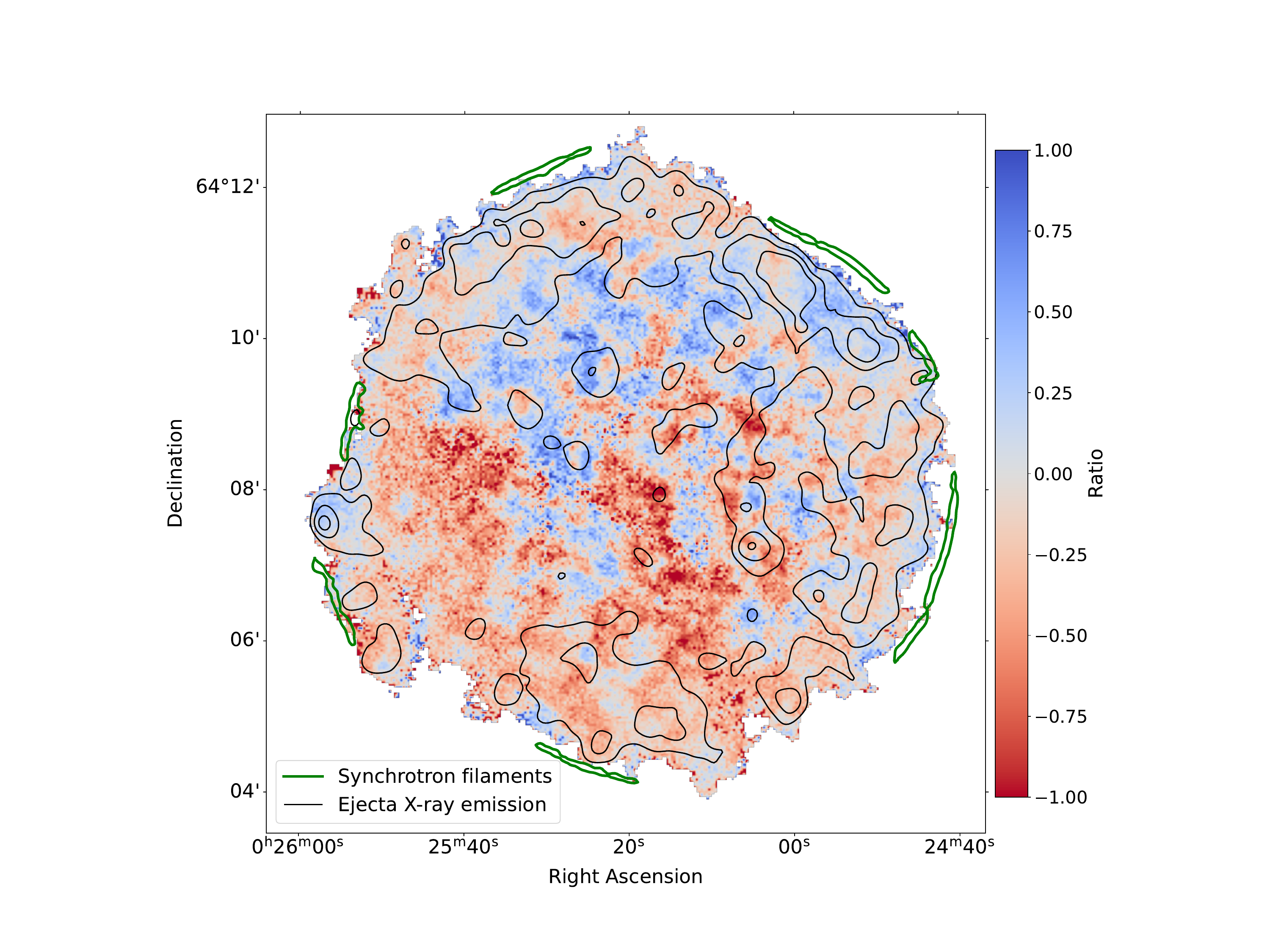}
\caption{\footnotesize Ratio map (see Eq. \ref{EqRatio}) of the red and blueshifted GMCA images. For example, a pixel with a value of 1 is dominated by the blueshifted GMCA image. The main synchrotron filaments are indicated in green, found with a contour detector in the first GMCA component. The dark contours come from the total image of the SNR in the 1.6 - 2.1 keV band smoothed by a 5 arcsec Gaussian kernel.}
\label{CarteRatio}
\end{figure*}

\section{Results: LoS velocities, $V_{\rm z}$}
\label{Section:Line of sight velocities}

In principle, we expect the X-ray emission from Tycho's SNR to arise approximately from a shell with half going toward us (blueshifted emission) and half away from us (redshifted emission). What we see in a pixel is the sum of emissions in the LoS because the SNR X-ray emission is optically thin. In the outputs found by GMCA (see Section \ref{subsection:General Morphological Component Analysis tool}) in Fig. \ref{FigGMCA}, two components are associated with ejecta emission. The major difference being the shift of the silicon line in their spectra, these components are interpreted as redshifted and blueshifted emission of Si ejecta.

To determine where the Si ejecta emission is predominantly blue or redshifted, we produced their ratio map from the GMCA outputs. Then we explored the possibility to derive physical maps from these two components: a map of Si line peak energy ($E_{\rm p}$) and a map of the $V_{\rm z}$.

\subsection{Ratio map of the red and blueshifted emission}

As explained previously, each component is optimized to reproduce best the true spectrum in a pixel as a linear combination of the GMCA spectra. The weight of each spectrum of the GMCA basis is the value in each pixel of the GMCA image. To investigate the dominance of one ejecta component against the other and highlight some asymmetries, we computed a ratio of the GMCA images which is defined as:

\begin{equation}
    \rm{Map} = \frac{\rm{S}_{\rm{blue}} - \rm{S}_{\rm{red}}}{\rm{S}_{\rm{blue}} + \rm{S}_{\rm{red}}}
    \label{EqRatio}
.\end{equation}

So, this ratio map can be expressed as follows: a pixel with a ratio of $-1$ is dominated by redshifted emission while a ratio of 1 is a dominantly blueshifted and zero if both component are  equal.
The resulting map is shown in Fig. \ref{CarteRatio}.
The lack of strong correlation between the red and blue structures and the brightness contours show that our map is independent of brightness.

We also observe  a clear asymmetry in the map with the south more redshifted than the north, but it is difficult to do more interpretation without a physical meaning for the ratio of components. Thus, we need to construct a physical proxy of the $V_{\rm z}$ measurement.

\subsection{From ratio map to velocity map}
\label{subsection:ratiotovelocity}

In this section, we explore how our red and blueshifted maps can be used as a proxy to estimate the mean $V_{\rm z}$ velocity in each pixel.
According to the GMCA definition, the spectrum in a pixel ($i,j$),  $\rm{A_{\emph{ij},\ tot}(\emph{E})}$ can be written as:

\begin{equation}
    \rm{A_{\emph{ij},\ tot}(\emph{E}) = \sum_{\emph{k}} S_{\emph{ij},\ \emph{k}}\ A_{\emph{k},\ GMCA}(\emph{E}),}
\end{equation}

with $\rm{S_{\emph{ij,\ k}}}$  as the value of pixel ($i,j$) in the image of the kth component and $\rm{A_{\emph{k},\ GMCA}(\emph{E})}$ as its spectrum. As we study the ejecta dynamics, we discarded the synchrotron component and we considered only the two line components. We can approximate their GMCA spectra as Gaussian functions as in the following equation (Eq. \ref{eqGaussienne}):
\begin{equation}
    \rm{A_{\emph{k},\ GMCA}(\emph{E}) = \beta_\emph{k} + \alpha_\emph{k} \exp{  \left(  - \left(\frac{ \emph{E} - \Bar{\emph{E}}_\emph{k}}{\sigma_\emph{k} }\right)^2  \right) }}
    \label{eqGaussienne}
,\end{equation}

If we want to find the peak energy of the silicon line in a pixel ($i,j$), the energy where the line reaches its maximum value, we must find the value of $E_{\rm{p},\  \emph{ij}}$, such that 

\begin{multline}
 \frac{d A_{\emph{ij}, tot}}{d\emph{E}}\big\rvert_{\emph{E}_{p,\ \emph{ij}}} = 0 \Leftrightarrow \\
\sum_{\emph k = 2,3} S_{\emph{ij,\ k}}\ \alpha_{\emph k}\ \frac{ - 2\ (\emph{E}_{p,\ ij} - \Bar{\emph E}_{\emph k})}{\sigma_{\emph k}^2}\  \exp{\left( - \left(\frac{ \emph{E}_{p,\ ij} - \Bar{\emph E}_{\emph k}}{\sigma_{\emph k}}\right)^2 \right) }  = 0. 
\end{multline}

If we do a limited development on the order of 1 (we suppose that $\rm{\mid E_p- \Bar{E}_k \mid \ \ll \sigma_k}$), we obtain Eq. \ref{EquCarteEc} to solve and its solution for two components: 

\begin{equation}
\rm{
    \sum_{\emph k=2,3} S_{\emph{ij,\ k}}\ \alpha_{\emph k}\ \frac{ (\emph{E}_{p,\ \emph{ij}} - \Bar{\emph E}_{\emph k})}{\sigma_{\emph k}^2}\ = 0 \Leftrightarrow \emph{E}_{p,\ \emph{ij}} = \frac{S_{\emph{ij},r} \frac{\alpha_r}{\sigma_r^2} \Bar{\emph E}_r + S_{\emph{ij},b} \frac{\alpha_b}{\sigma_b^2} \Bar{\emph E}_b }
{S_{\emph{ij},r} \frac{\alpha_r}{\sigma_r^2} + S_{\emph{ij},b} \frac{\alpha_b}{\sigma_b^2} } }
\label{EquCarteEc}
.\end{equation}

In our case, $E_{\rm r}$ and $E_{\rm b}$ are the red and blueshifted peak energies of the two GMCA spectra of ejecta, $\rm{\sigma_r}$ and $\rm{\sigma_b}$ as their widths, and $\rm{\alpha_r}$ and $\rm{\alpha_b}$ as their normalizations.
\\

\begin{figure}
\centering
\includegraphics[scale=0.45, trim = 0 0 0 0, clip=true]{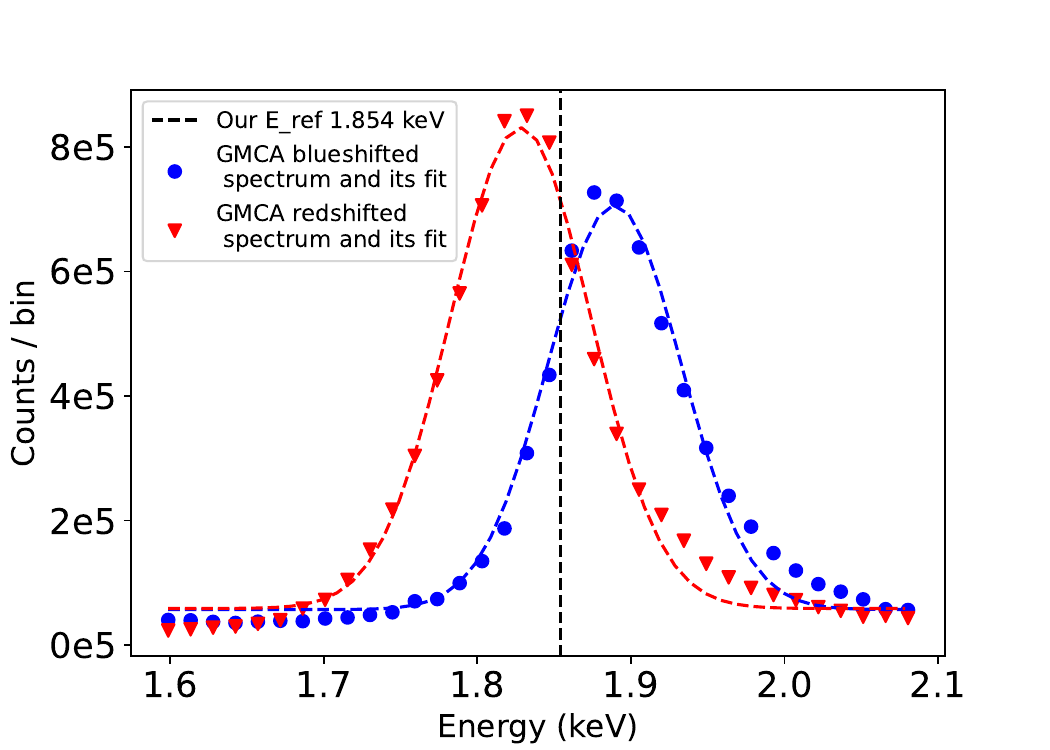}
\caption{\footnotesize Gaussian fits of the red and blueshifted GMCA spectra. There are no uncertainties in the GMCA outputs.}
\label{fitSpectreGMCA}
\end{figure}

We apply this protocol to the output of GMCA, as shown in Fig. \ref{FigGMCA}. The fitting of the GMCA spectra shown in Fig. \ref{fitSpectreGMCA} gives the values from Table \ref{table:2}. It provides a reasonable fit to the data. The last row in the table is the velocity corresponding to the peak energy of the GMCA spectra for the energy of reference of 1.854 keV (see Appendix \ref{Appendix : Energy at rest }).  

\begin{figure*}
\centering
\includegraphics[scale=0.4, trim = 50 70 0 100, clip=true]{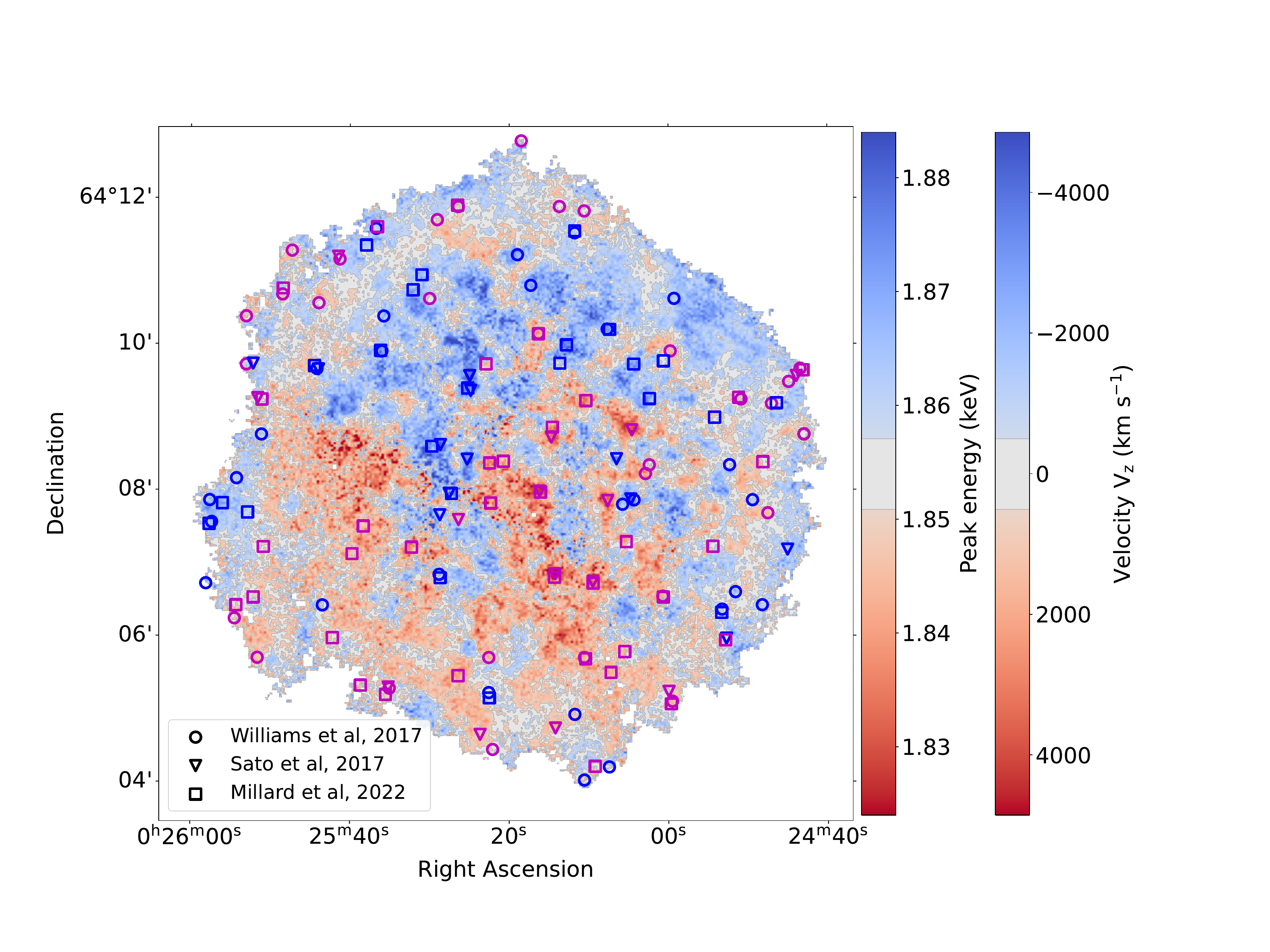}
\caption{\footnotesize Map of the peak energy reconstructed with GMCA components and its equivalence in terms of integrated velocity in the LoS. The markers are the measurements from \cite{Williams2017} shown as circles,\cite{Sato2017a}, shown as triangles, and \cite{Millard2022} shown as squares. Their colors (blueshifted or redshifted) come from their results.}
\label{CarteEc}
\end{figure*}

\begin{table}[h!]
\caption{\footnotesize GMCA spectral fit with Gaussian functions}             
\label{table:2}      
\centering                          
\begin{tabular}{c c c}        
\hline\hline                 
Spectral parameters & Redshifted & Blueshifted \\    
\hline                        
   $\beta_k$ (Counts/bin) & 5.9$\times10^4$  &  5.7$\times10^4$ \\      
   $\alpha_k$ (Counts/bin) & 7.7$\times10^5$ & 6.5$\times10^5$   \\
   $\sigma_k$ (keV) & 0.0646  & 0.0615    \\
   $\Bar{E}_{ k}$ (keV) & 1.8282   & 1.8894   \\
   $V_{\rm z}$ associated to $\Bar{E}_{ k}$ (km s$^{-1}$)  & -4175 & 5722    \\ 
\hline                                   
\end{tabular}
\end{table}

Then we use Eq. \ref{EquCarteEc}, which can be applied directly on the GMCA images $S_{\rm r}$ and $S_{\rm b}$, to obtain a map of a proxy of the peak energy of the silicon line in all the SNR, which is shown in Fig. \ref{CarteEc}. With a simple transformation, we can obtain $V_{\rm z}$: \( V_{\rm z} = \frac{E_{\rm p} - E_{\rm ref}}{E_{\rm ref}}\ c\). We suppose that the energy of reference of the silicon line, which means the energy without the Doppler effect due to expansion, is identical in all the SNR. This energy at rest depends on the local temperature and ionization state, so this is an important question that we study more in detail in Appendix \ref{Appendix : Energy at rest }. With our investigation, we take a value of 1.854 keV for all the SNR.
We note that while we find it reasonable for the case of Tycho, this is not a valid assumption in general and it might even be a limitation in applying this method to other objects.

Figure \ref{CarteEc} is very similar to the ratio map (see Fig. \ref{CarteRatio}). We did not add again the synchrotron filaments and the X-ray emission contours, but the conclusion is the same, our results are brightness-independent. Concerning systematic uncertainties, the discussion in Appendix \ref{Appendix : Energy at rest } about the energy of reference and discussion in \cite{Williams2017} about the gain calibration of the \emph{Chandra} ACIS, it appears that systematic uncertainties will be of the order of $\pm$500 km s$^{-1}$ both for calibration and stability of the energy of reference across the SNR.
So in the colormap of Fig. \ref{CarteEc} we put a grey uniform color the velocity values between -500 km s$^{-1}$ and + 500 km s$^{-1}$.

\begin{figure}[t]
\centering
\includegraphics[scale=0.5, trim = 140 0 0 0, clip=true]{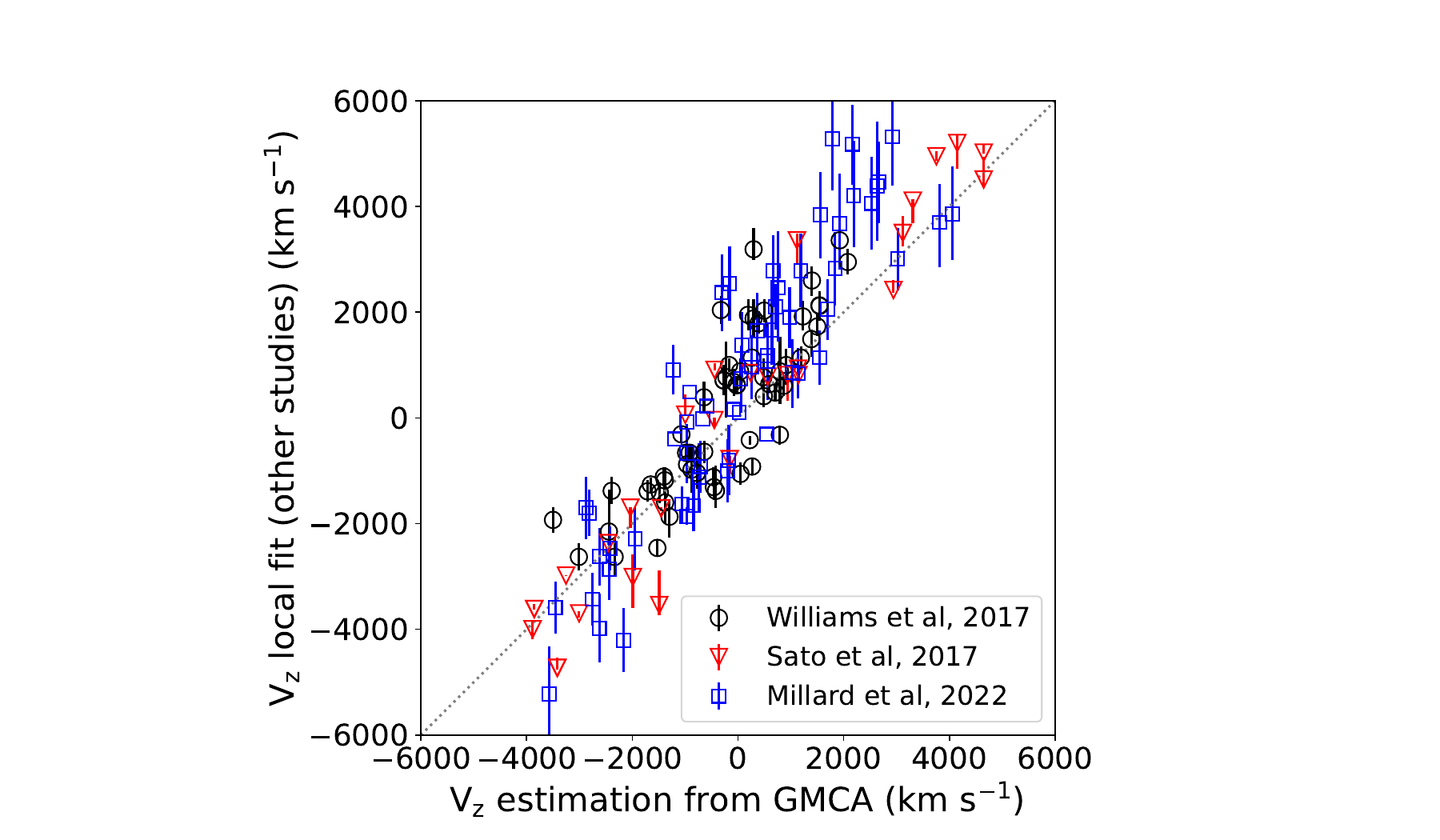}
\caption{\footnotesize Comparison of the $V_{\rm z}$ velocity obtained with our method and with spectral studies using ACIS (\cite{Williams2017} with black markers and \cite{Sato2017a} with red markers, and gratings \citep{Millard2022} with blue markers.}
\label{ComparaisonArticle}
\end{figure}

For comparison we show in Fig. \ref{CarteEc} the zones of extraction and $V_{\rm z}$ results of spectral studies using the ACIS CCD \citep{Williams2017, Sato2017a} and the HETG gratings \citep{Millard2022}.
The first observation is that our map seems to match locally their values. We cross-check our method in Fig. \ref{ComparaisonArticle} by comparing their velocity estimate derived from the spectral fitting with our GMCA velocity estimate taken at their extraction region position. In conclusion, our global method is in very good agreement with their local results with some dispersion that might be due to variations in the plasma temperature and ionization timescale that will modify the Silicon energy of reference.

We can easily explain why this map and the ratio map \( \rm{ \frac{\emph{S}_{red} - \emph{S}_{blue}}{\emph{S}_{red} + \emph{S}_{blue}} }\) are nearly the same. If we suppose that the two GMCA spectra are the same spectrum but just translated, that means \(\rm{\alpha_r \approx \alpha_b}\) and \(\rm{\sigma_r \approx \sigma_b}\). With the hypothesis that the shifts are symmetrical, such that \(\rm{ \Bar{\emph{E}}_r = \emph{E}_{ref} - \Delta \emph{E}}\) and \(\rm{ \Bar{\emph{E}}_b = \emph{E}_{ref} + \Delta \emph{E}}\), Eq. \ref{EquCarteEc} becomes:

\begin{equation}
  \rm{\emph{E}_P = \emph{E}_{ref} + \Delta \emph{E} \ \frac{\emph{S}_{red} - \emph{S}_{blue}}{\emph{S}_{red} + \emph{S}_{blue}} }
.\end{equation}

This map is proportional to the ratio map (Fig. \ref{CarteRatio}). We see in Table \ref{table:2} and Fig. \ref{ComparaisonArticle} that all these hypotheses are nearly valid. We remark that the velocities on the LoS associated to the red and blueshifted GMCA spectra are not symmetrical according to zero. This is why a simple ratio as we did in the first approach is not enough: the zero in Fig. \ref{CarteRatio} does not correspond to a null redshift. We can see Eq. \ref{EquCarteEc} as a weighted mean of the GMCA images with the GMCA spectral parameters correcting this effect. 

Thus, our developments to have a physical map are justified, but we must keep in mind limitations of our method, namely: 1) the GMCA tool does not contain any management of uncertainties, neither for images nor spectra; 2) the choice of this Gaussian function may not provide the best fitting for the GMCA spectrum. However, it is the easiest way to obtain the peak energy map analytically; 3) the most limiting approximation is probably the simplification of the exponential to find an easy analytic solution for the map of peak energy. So our map must be seen as a proxy of the integrated peak energy and $V_{\rm z}$. 4) Finally, the transformation from peak energy to $V_{\rm z}$ raises the question of the energy of reference, which is studied in more depth in Appendix \ref{Appendix : Energy at rest}. It is only an average velocity weighted by the local flux in the line of sight is reconstructed.

In conclusion, we obtain a map of the mean $V_{\rm z}$ at a 2" pixel level with a total coverage of the SNR for the first time. Overall we find higher $V_{\rm z}$ in the center than at the edge, as expected for a spherical expansion but with many patchy features dominating in the foreground or in the backgroun.
The important features which will be discussed in more detail later is the clear north and south asymmetry. A similar trend was seen by \cite{Millard2022}, but with limited sampling and is clearly confirmed here thanks to our full coverage.

\subsection{Limitations due to integration in the LoS}
\label{Limitations due to integration on the line of sight}

As SNRs are optically thin in the X-ray domain, there is a notable difficulty with the map of $V_{\rm z}$ (Fig. \ref{CarteEc}). The spectrum in a pixel is integrated over the LoS, so the peak energy and the corresponding velocity are also weighted by the local brightness along the LoS. 

In a perfectly spherical remnant with a homogeneous emission and spherical expansion, velocities from the two half-shells would cancel each other out and no Doppler shift would be measured with our method, with only a line broadening observed. However, even for a regular type Ia supernova remnant, these assumptions are not valid. In particular, the flux varies at large scale in the SNR. Due to the Rayleigh-Taylor instabilities and the clumpy aspect of the ejecta, bright clumps can also dominate the LoS.

This is why the usual method used in \cite{Williams2017}, \cite{Sato2017a}, and \cite{Millard2022}, which consists of studying only bright blobs is interesting. By isotropy, we can suppose that the blob is also locally small in the LoS and that it dominates the emission. In this case, the measures of $V_{\rm z}$  are located at one point on the LoS (which is necessary for a 3D reconstruction). But it is difficult to find good blobs to study in the SNR. There are around a hundred points currently in the accumulation of all the studies using this method, providing a limited coverage of the SNR.

Our method is complementary: we can quickly obtain a proxy of the integrated $V_{\rm z}$  with total coverage. We are consistent with the local measurements (see Fig. \ref{ComparaisonArticle}) and we highlight some large-scale asymmetries (see the discussion in Section \ref{largescaleVz}). However, there are some limitations with our method. First, we have a degeneracy at low values (in grey in Fig. \ref{CarteEc}): it can be due to true slow velocities (which are expected at the edge), or compensation of the two SNR halved. Then, our values tend to be underestimated (the slope of the correlation in Fig. \ref{ComparaisonArticle} is greater than 1). In general, there probably is no clear dominance of one side over another. Thus, the velocities are averaged on the LoS. Then, we also have a problem in terms of the interpretation of these large-scale asymetries: both the asymmetries of velocity and flux  may explain them. Finally, it is difficult to localize the position of the emitting region in the LoS to do a full 3D reconstruction (as we present in Section \ref{subsection:3D}).


\section{Results : Plane-of-the sky,  $V_{\rm xy}$}
\label{Section:Plane of the sky velocities}

\begin{figure*}
\centering
\includegraphics[scale=0.4, trim = 50 0 0 50, clip=true]{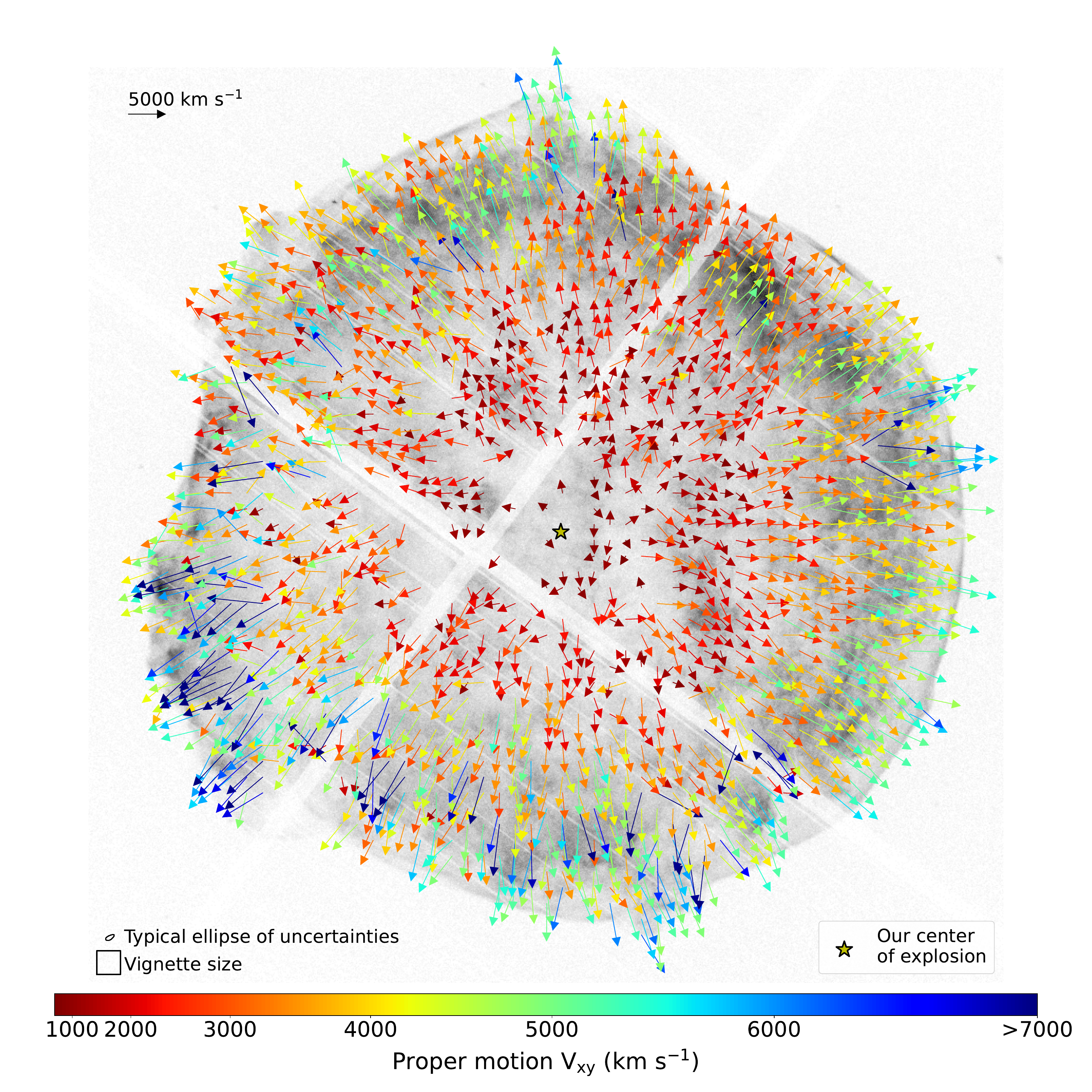}
\caption{\footnotesize Proper motion vector field between 2003 and 2009 obtained with the tool POF. The size of the vignette for each tracked feature is 15". There are 1722 vectors, colored by the value of their norm. The colorbar is saturated at 7000 km s$^{-1}$. The background image is the observation of Tycho's SNR in 2003 without any exposure map correction.}
\label{SOF}
\end{figure*}

\subsection{Proper-motion vector field}
\label{subsection : The vector field}

As explained in Section \ref{subsection:Optical Flow to measure proper motion}, we used POF, a tool we devised to compute around 1700 proper motion vectors in the plane of the sky. The first step of our method is to find small morphological features whose shift between two observations at different epochs will be measured.

We initially tested the method only on bright features with a sharp morphology, such as ejecta knots. However, it appeared that when extending the method to fainter less contrasted features, the tool succeeded to measure a shift with reasonable likelihood profiles providing relatively small errors in all directions (as shown in Appendix \ref{Appendix : POF}, see Fig. \ref{fig:profilPOF}, center panel).
Our method is not only sensitive to bright knots, but also to more diffuse structures. This is because we are using the full 2D information in the likelihood and not only a 1D projection, which necessarily produces some loss of information.
Following these tests, we decided to not only follow bright knots but map the proper motion of the entire remnant by tracking features defined on a regular grid of points. We take only the points inside a mask of ejecta created with the GMCA outputs, excluding most of the synchrotron filaments. The boxes around the good features (called vignettes in the following) are 30 pixels (15") wide and their centers are separated by 20 pixels. So, there is some overlap from one box to another.

Then we applied POF to the epochs 2003 and 2009. The 2009 observation has a deeper exposure time and will be used as the model. The maximal shift that can be measured in the exploration zone is 8 pixels (corresponding to $\sim$11000 km s$^{-1}$).

We must then deal with the anomalies in our outputs. If an initial feature is located in a zone without enough counts and/or contrast, our tool will not succeed in measuring a shift. Most of the vignettes where the method provides unreliable results are located close to the exposure gaps in the 2003 observation (i.e., on the bad columns and the CCD gaps). 
Thus, we chose to keep only the measurements with an expansion index, $m$ (where $m = \frac{V_{\rm xy} t_{\rm SNR}}{R_{\rm xy}}$), that is lower than 1.2 (some of them were up to 12). This removes around 8.6 \% of the outputs, essentially in the areas of the CCD gaps. 

For the parameters described above, we obtained 1722 velocity vectors (shown in Fig. \ref{SOF}) out of 1884 initial features in the grid. The ellipses of uncertainty, different for each vector, are not shown here, for more readability. The mean value for these 1-sigma uncertainties on the vector direction is 370 km s$^{-1}$ (see Fig. \ref{fig:profilPOF} for some examples).  

As expected for a nearly spherical expansion projected on the plane of the sky, the velocity in the plane of the sky is higher at the edge than in the center. The distribution of $V_{\rm xy}$ has a mean value of around 3610 km s$^{-1}$ or 0".217 yr$^{-1}$. This corresponds to a mean expansion index of 0.59. Our values are consistent with previous studies about ejecta dynamics: \cite{Williams2017} found a mean $V_{\rm xy}$ velocity of 4430 km s$^{-1}$ with a range from 2400 to 6600 km s$^{-1}$ and \cite{Millard2022} reported a mean of 4150 km s$^{-1}$ in the range from 1890 to 5950 km s$^{-1}$.

Our vector field in Fig. \ref{SOF} may seem a bit noisy, as the vectors are not perfectly radial in general. There is no spatial regularisation, so we probed only the local behavior. Physically, they can be non-radial because of local turbulence and deviations due to interactions with dense clumps or large interstellar clouds. 
The angular deviation distribution between our vectors and a radial equivalent is a Gaussian centered at 3.1° with a standard deviation of 19.6°.

\subsection{Center of the explosion from the vector field}
\label{subsection : Center of the explosion from the vector field}

\begin{figure}[h!]
\centering
\includegraphics[scale=0.35, trim = 5 0 0 0, clip=true]{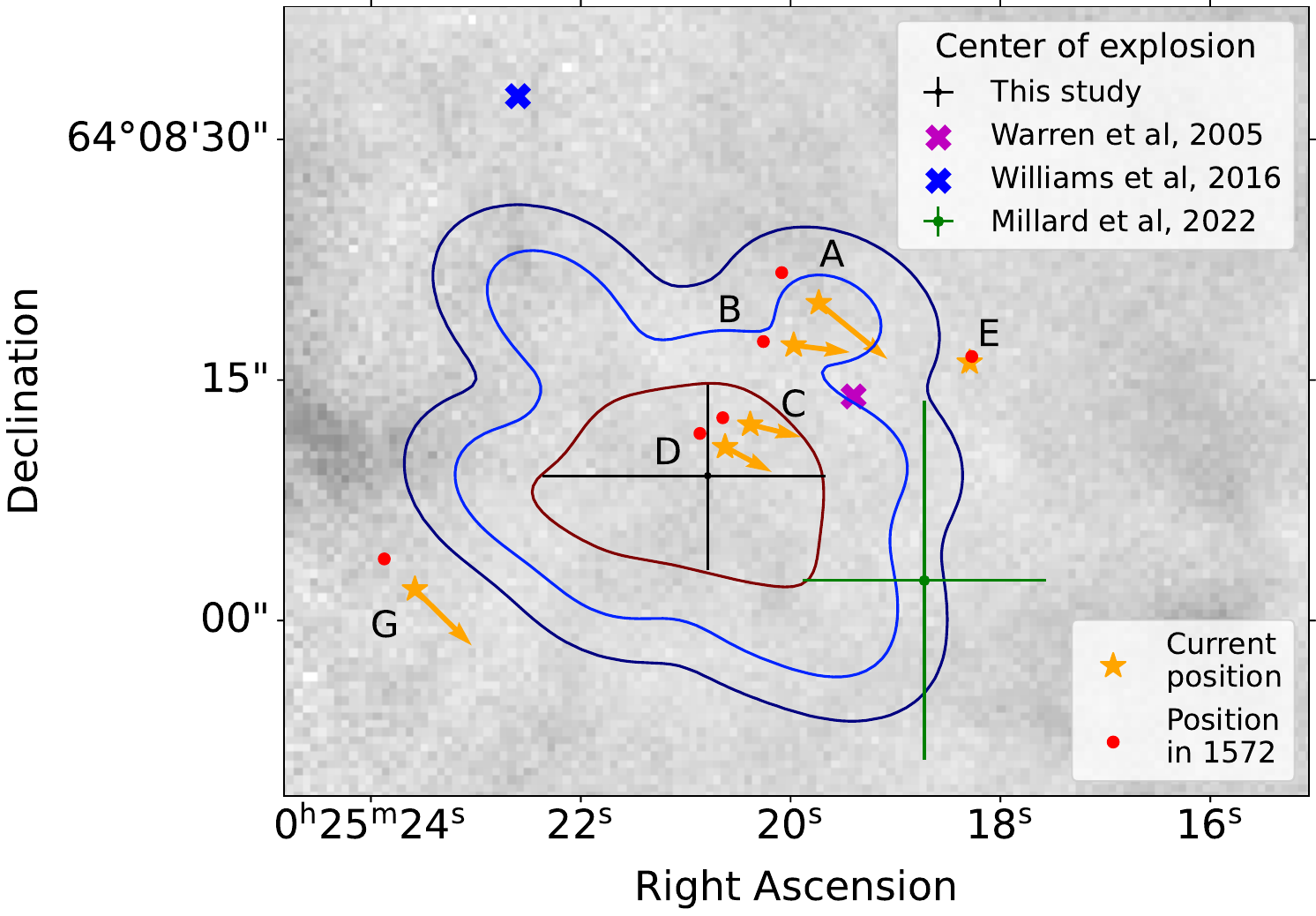}
\caption{\footnotesize Locations of the center of the explosion found by this study (see main text) and by other studies. The contours are our uncertainties at 1, 2, and 3 sigma based on Gaussian kernel density estimation. The star symbols are the potential donors listed by \cite{Kerzendorf2013} together with their current and past positions and proper motion vectors from \emph{Gaia} DR3.}
\label{CentreExplosion}
\end{figure}

With this velocity vector field, we can attempt to find the common origin of these vectors, which is the center of the explosion. To do this, we used the method from \cite{Sato2017b}. The idea is to suppose a power-law radius expansion $ r \propto t^m$, where $m$ is the expansion index. If $m$ is low, the ejecta have slowed down; then, if $m$ is near 1, the ejecta are in free expansion.
Under these assumptions, the projected radius, $R_{\rm xy}$, is equal to  $ \frac{V_{\rm xy} t_{\rm SNR}}{m}$. And so, the center of the explosion can be deduced from the position of each vector, its expansion index, and the age of the SNR, $t_{\rm SNR}$. However, the origin of the explosion is also needed to calculate $m$. 
To solve this, we use an iterative protocol. We initiate it with the center from \cite{Williams2016} and proceed as follows:

First, we calculate $m$ and the angular deviation $\Delta \theta$ from a pure radial expansion for each vector. Second, we create a mask to have a "golden sample" of vectors with $1.2>m>m_{\rm lim} $ and $\vert \Delta \theta \vert < \Delta \theta_{\rm lim} $, to keep vectors that have not decelerated too much and have had little angular deviation. Third, we calculate an origin for each vector of this golden sample, which is in the direction of the vector at a distance of $R_{\rm xy} = \frac{V_{\rm xy} t_{\rm SNR}}{m}$.
Finally, we take the median of these origins as a new center.

We used the distribution of origins at the final step to obtain the uncertainty contours using a Gaussian kernel density estimate. The final value and its error bar are the median and standard deviation of this last distribution, as shown in Fig. \ref{CentreExplosion}, together with a comparison with previous studies. We used a hundred iterations and, in practice, the convergence is very fast. As a limit for the golden sample, we took a maximal deviation from radial vector  of $\Delta \theta_{lim}$ = 5° and a minimal expansion index of  $m_{lim} = 0.75$. Finally, 44 vectors remained, which is many more than all the other studies using this method; we also obtained a value of R.A. = 00$^h$25$^m$20$^s$.79 $^{+12.3"}_{-10.3"}$ and Dec = 64°08'09".04  $^{+5.7"}_{-5.9"}$. 

Our result is closer to the measurement of \cite{Warren2005}, which was based on geometrical considerations. The measurement from \cite{Williams2016} that we have used as the starting point is based on the measurement of forward shock expansion in 17 regions and a relation to measure center of the explosion offset from geometrical center based on simulation  \citep{Williams2013}. Our result uses the ejecta as tracer, which is more directly connected to the explosion than the forward shock. The latter is more sensitive to the circumstellar medium and the perturbations due to the expansion. The result from \cite{Millard2022}, which uses the same protocol but with fewer vectors, is also compatible with our result. Given this new estimate of the explosion center, we carried out a search for a potential progenitor using \emph{Gaia} Data Release 3 (DR3), presented in Section \ref{gaia}.


\section{Discussion}
\label{Section:Discussion}

\begin{figure}
\centering
\includegraphics[scale=0.4, trim = 00 0 0 0, clip=true]{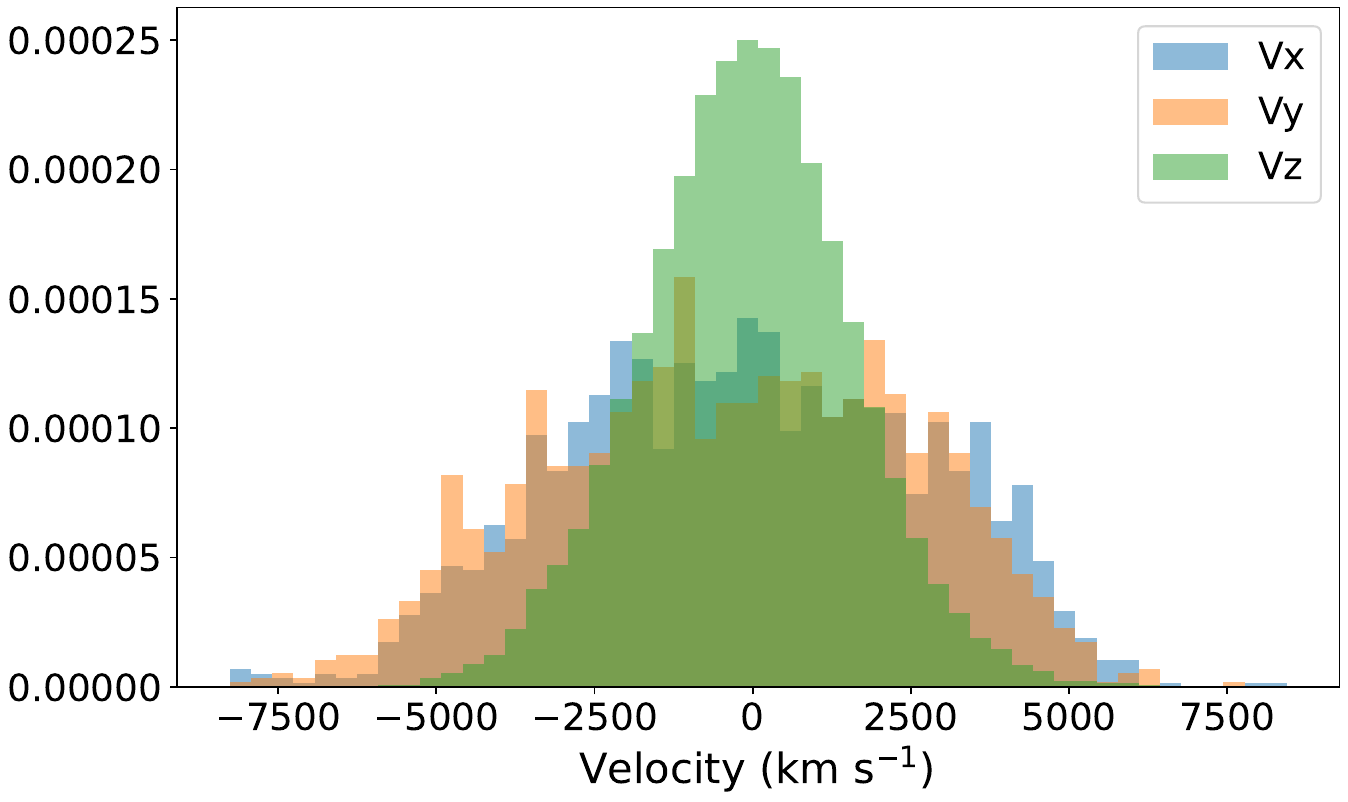}
\caption{\footnotesize Normalized histograms of our velocities,$V_{\rm x}$ and $V_{\rm y}$, obtained with the proper motion method and $V_{\rm z}$ estimated via the Doppler shift. The x-axis is oriented to the west, the y-axis to the north, and the z-axis is positive away from the observer.}
\label{histVxyz}
\end{figure}

Thanks to the two new velocity measurement methods that we developed in this study, we obtained around 1700 $V_{\rm xy}$ vectors and a $V_{\rm z}$ map, at a 2'' spatial resolution, with a total coverage of the SNR. All our measurements are summarized in the histogram in Fig. \ref{histVxyz}. As explained in Section \ref{Limitations due to integration on the line of sight}, our distribution of $V_{\rm z}$ is biased by the integration along the LoS and the velocities are likely to be underestimated for high values.
At first glance, the velocities on the three axes are in a range between -6000 and 6000 km s$^{-1}$ with a symmetric distribution. As expected for a SNR issued from a thermonuclear supernova, Tycho's SNR has a regular shape and dynamics overall.

However a detailed inspection reveals a more complex behavior with dynamics asymmetries both at large and small scales. The origin of these behaviors can be innate, which means due to the explosion anisotropy, or acquired because of inhomogeneities in the environment that slow down the expansion. In this second case the question is also raised to know the age of this interaction and the origin of this density inohomogeneities: is it due to the progenitor (circumstellar medium, CSM) or was it pre-existing (interstellar medium, ISM)?

In the following sections (Sections \ref{largescaleVz}, \ref{largescaleVxy}, and \ref{smallscale}), velocity maps are used to investigate large and small scales dynamics anisotropies in the context of our understanding of the surrounding medium. Then we discuss the use of the ejecta vector field to pinpoint the explosion center and search for stellar progenitors (Section \ref{gaia}). Finally we combine all the velocity information into a 3D representation in Section \ref{subsection:3D}.

\begin{figure*}
\centering
\includegraphics[scale=0.5, trim = 25 80 20 120, clip=true]{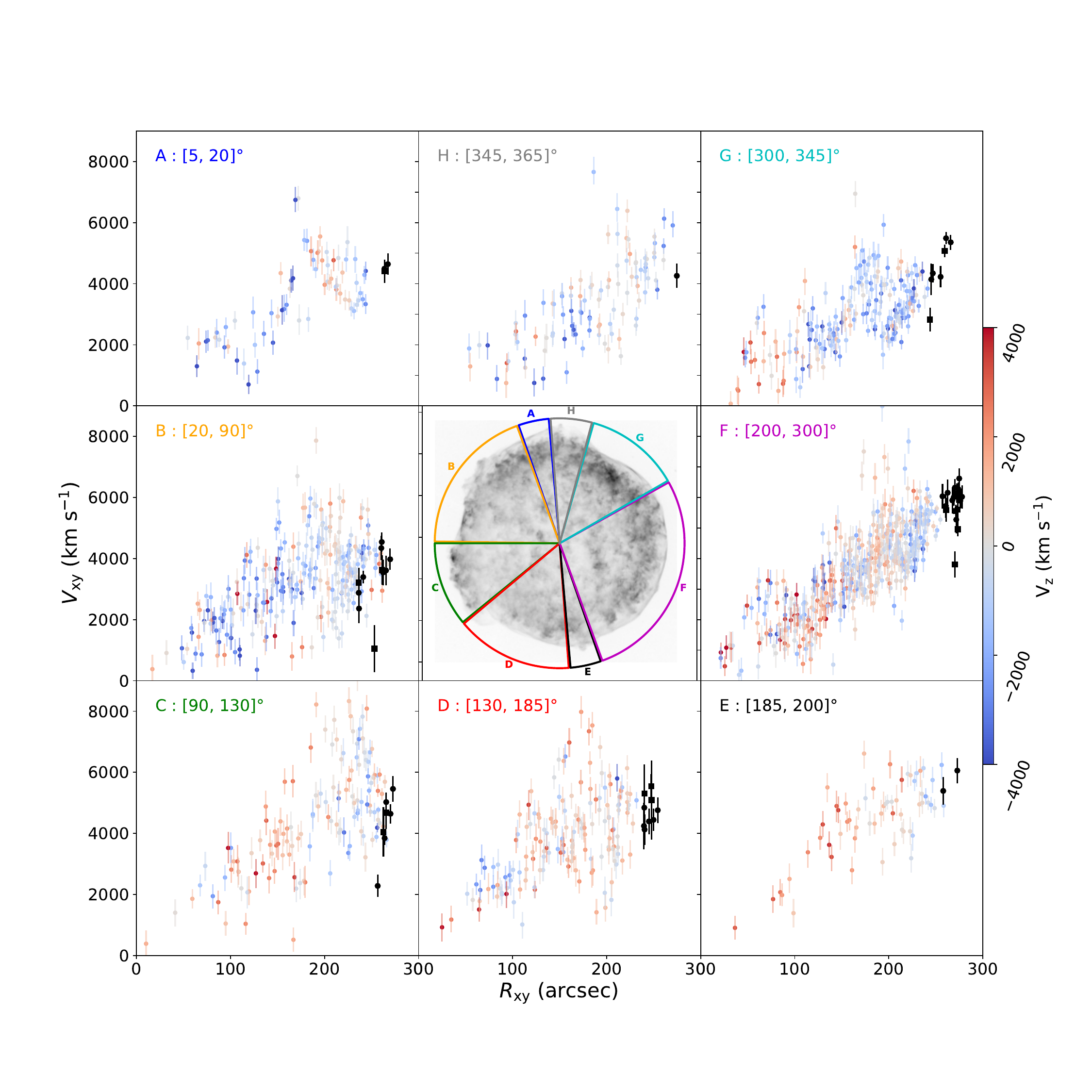}
\caption{\footnotesize Profiles of the proper motion, $V_{\rm xy}$, as a function of the radius in the plane of the sky, $R_{\rm xy}$, for eight angular sectors. In the central panel, sectors are overlaid on the 0.5-7 keV \emph{Chandra} map from the deep 2009 observations. The color red/blue is the velocity $V_{\rm z}$ of our map in Fig. \ref{CarteEc}, in the same position as the POF measurements in Fig. \ref{SOF}. We add on the profiles the forward shock velocity measurements from \cite{Katsuda2010} in black circles and \cite{Williams2016} in black squares that are located in the associated sector.}
\label{polarSOF}
\end{figure*}

\subsection{Large-scale asymmetries in the $V_{\rm z}$ map}
\label{largescaleVz}

The most obvious large-scale asymmetry is in the LoS velocity $V_{\rm z}$ from Fig. \ref{CarteEc}. In the south, the redshifted emission is dominant and in the north, there is more blueshifted emission. This point has been also noted in \cite{Millard2022} and \cite{Sato2017a}, but with a much more limited coverage. This is clearly confirmed in our work, with our full coverage of the entire remnant. This means that in the north, the bulk of the material is preferentially moving toward us or that the near side is brighter than the back side. In the south, it is the opposite.

There are two ways to interpret this asymmetry. This can be due to an asymmetry during the explosion. Hundreds of years after, the explosion asymmetry could still be visible in the SNR structure, as shown, for example, in the type Ia simulation of \cite{Ferrand2019}. Perhaps the asymmetry that we see is due to the SNR being an oblong shell, elongated at an angle with respect to the LoS creating this blue and red patterns in the north and in the south, respectively.

Another possibility is that this asymmetry is acquired due to an interaction with interstellar material that slows down the ejecta. If a cloud is behind the SNR in the north and another in front of the SNR in the south, we would have the type of ejecta dynamics that we observe. However, these interactions would increase the brightness of the slower side of the shell. As we measure the integrated velocity along the LoS, this could compensate the higher velocity of the non-interacting side.
A ring of circumstellar matter could explain this distribution as is observed for SN 1987a. Nevertheless, there are no clear observations of a cloud in front of the SNR in the south. This hypothesis of a slow-down due to an interaction raises the question of when such an interaction could have happened. 
Either the SNR is currently interacting with large scale clouds or this asymmetry was acquired during the first decades after the explosion in a scenario in which the SNR had evolved in a dense, but small, wind bubble as described in \cite{Chiotellis2013}.
The second possibility could explain why we do not currently observe a cloud in front of the southern half. Nevertheless, in their 1D simulation, they show that current dynamics of Tycho's SNR will be identical to a case without a wind shell; namely:\ there will be only an impact on the ionization time and a small variation of the reverse shock's radius.
Subsequent works could disentangle these scenarios with a complete mapping of the ejecta plasma parameters (in particular the plasma temperature and ionization timescale) via X-ray spectral analysis.

\subsection{Large-scale asymmetries on the $V_{\rm xy}$ vector field}
\label{largescaleVxy}

A large-scale asymmetry is also seen in the plane of the sky. This was noticed in forward-shock proper-motion measurements from \cite{Williams2016}, \cite{Katsuda2010}, and \cite{Tanaka2021}. In \cite{Williams2013}, mid-infrared observations highlighted a density gradient from east to west, which agrees with the forward shock asymmetries. This east-to-west asymmetry was not observed in the ejecta proper motion vector field \citep{Williams2017, Millard2022}.

At first glance, in our vector field in Fig. \ref{SOF}, it is difficult to say if the ejecta also show this east-west asymmetry. To study  our proper-motion vector field in more detail, we represent in Fig. \ref{polarSOF}, the profiles of the $V_{\rm xy}$  as a function of the radius in the plane of the sky, $R_{\rm xy}$, for eight angular sectors. These sectors are based on the morphology of Tycho's SNR, as seen in the middle panel and details in its proper motion's dynamics. Sector F between 200 and 300 degrees is used as a reference of an expected dynamics without perturbations in other analyses \citep{Chiotellis2013, Badenes2006}.
Sector C matches the known fast iron and silicon rich knots \citep{Yamaguchi2017}. Sectors E and H represent the protrusions, where the ejecta reach the forward shock.
The small sector A was selected because of its unexpected dynamics seen in Fig. \ref{SOF}. Sector G has higher flux and slower velocities at the edge. Finally, sectors B and D probe large scale dynamics in the east and south-east, where the forward shock is slower. In this figure, we also added the forward shock measurements from \cite{Katsuda2010} and \cite{Williams2016}, based on \emph{Chandra} data using the same 3.5 kpc distance as in this paper.

We can see the forward shock asymmetry with velocities up to 6000 km s$^{-1}$ in the west (sector F) and slower velocities of around 4000 km s$^{-1}$ or less in the east (sector B). The initial observation is that we do not see the same contrast in our ejecta velocities. 
However, when comparing the ejecta dynamics with forward shock dynamics, there is a clear pattern where in the west, the forward shock moves (as expected) faster than the ejected material. Due to projection effects, there is also a linear relation between $V_{\rm xy}$ and $R_{\rm xy}$, with very few deviations for the sector F.  In the east, in sectors B and D, for example, the forward shock is slower than the ejecta and we observe strong variations around the expected linear behaviour.

Maps of the ambient medium at the edge of the SNR based on infrared \citep{Williams2013} and radio observations \citep{Arias2019, Castelletti2021} show that the west has indeed no potential clouds that could potentially disturb the spherical expansion. Nevertheless, \cite{Chiotellis2013} argued that an interaction with a small and dense wind bubble during the early expansion phase (less than 100 years) of Tycho's SNR could explain the dynamics and spectral properties in the sector F.
The east, on the contrary, seems to have a complex external structure in the multi-wavelength observations, which explains the velocity difference between the forward shock and the ejecta. This medium could also be the origin of some local anomalies in our vector field, as discussed in the next subsection.
Figure \ref{polarSOF} also includes the velocity $V_{\rm z}$ in the color of each marker. It shows that there are no correlations between the behaviour at large scale in the plane of the sky and in the LoS.

\subsection{Small-scale velocity anomalies on the edge of the SNR}
\label{smallscale}

Sector C at the south-east is known for its fast iron and silicon knots studied by \cite{Yamaguchi2017}. The same behavior is observed for Kepler's SNR by \cite{Sato2020}. These are visible in our vector field with speeds around 8000 km s$^{-1}$ (and even up to 9000 km s$^{-1}$, as shown in Figs. \ref{SOF} and \ref{polarSOF}). In their study of freely expanding knots in Kepler's SNR, \cite{Sato2017b} summarized two explanations for fast moving blobs: they can be formed in a high density contrast region or can propagate in a low-density ambient medium. Here, these knots are interpreted as clumps produced in the inner layers of the supernova and ejected at high speed during the explosion \citep{Wang2001}. Thus, their properties (temperature, ionization rate, etc.) as well as their dynamics are not similar to the surrounding ejecta.

Sectors E and H illustrate also protrusions where the ejecta reach the forward shock as in C. However, their profiles are more linear and there are no structures that differ from their environment in the morphology and in the $V_{\rm xy}$ vector field. It seems likely that these protrusions are due to a low-density ambient medium. In the map of radio optical-depth that traces the external densities from \cite{Arias2019}, these regions seem to coincide with low density windows or gaps between overdensities.

Sector A has no particular X-ray morphology as shown in the central panel of Fig. \ref{polarSOF} but is remarkable in the $V_{\rm xy}$ vector field. There is clearly a structure characterized by a high velocity at a lower radius contrary to the expected projected dynamics. Contrary to the other sectors, there is a correlation with the $V_{\rm z}$ velocity: the rising velocities in the profile are mainly blueshifted, the slow-down is mainly redshifted, and then the edge is blueshifted again. This suggests a complex structure of several fronts on the LoS.
A potential molecular cloud is known in this region \citep{Zhou2016} and the radio study also shows an enhancement of the density in this zone \citep{Arias2019, Castelletti2021}. This cloud seems to have a direct impact on the ejecta and not only on the forward shock, as it has been observed. 
We see two possibilities to explain this profile. The enhancement of the velocity could be a structure in the foreground, such as the fast knot in region C, but it is seen to be projected. In the background the ejecta could interact with the cloud, which would explain the slower forward shock. Or the profile could be seen as one structure entirely due to this current interaction with the cloud.

Due to the reflected shock, the ejecta would be slowed down over time. However in this case, it is surprising that the forward shock is not nearer the contact discontinuity. 
In both case a simulation, even 1D, which track the ejecta velocity according time and radius in case of an interaction, would be useful for interpreting our result.

Finally, section G is also perhaps a proof of interaction with an over-density in the ambient medium. The velocity on the edge is slower in comparison to the other sectors,  around 4000 km s$^{-1}$. \cite{Williams2013} also showed an increase of the density in this region and its large flux could be due to this interaction.

\subsection{Searching for a surviving companion with \emph{Gaia} DR3 data}
\label{gaia}
With their new center of the explosion, slightly offset from previous studies, \cite{Williams2016} have argued that a new search of a potential progenitor could be done. However our new center of explosion is more compatible with the result found by \citet{Warren2005}, which was then used by \cite{Kerzendorf2013} in the optical search for companions.
The potential donor stars of \cite{Kerzendorf2013} are noted in Fig. \ref{CentreExplosion}. Overlaid we have the proper motion information from the \emph{Gaia} DR3 \citep{Gaia16,Gaia22} catalog and the position of each potential donor stars back in the year 1572 to be compared with our explosion center.

Candidates A, B, C, and D align with our center of the explosion, while candidates G and E are slightly outside our contours. The recent \emph{Gaia} DR3 release provides valuable updates on the parallaxes and distances associated with these objects\footnote{For reference the \emph{Gaia} DR3 source\_id are A: 431160565571641856, B:431160569875463936, C:431160363718444928, D:431160363709280768, E:431160565573859584, G:431160359413315328.}, which are summarized in Table \ref{tab:gaia}. Based on the derived geometrical distances from the parallax data, only stars B and E have distances in agreement with the SNR distance range of 2.5-4 kpc \citep[see Fig. 6 of][for a review]{Hayato2010} with distances of $2.53^{+0.23}_{-0.20}$ kpc and  $3.52^{+2.0}_{-1.0}$ kpc, respectively. Other candidate stars are within the 0-2 kpc range and are likely foreground stars.

Previously, the stellar candidate E was estimated to be at a distance of approximately 10 kpc \citep{Kerzendorf2013} or 7 kpc \citep{Ruiz-Lapuente2019}, making it unlikely to be associated with the SNR.
However, the latest parallax measurement from Gaia DR3 indicates that star E is much closer. The photo-geometric distance, a useful method for poorly measured parallax, takes into account both the parallax and photometric data to constrain the distance using stellar models. Using this method, the errors in the distance to star E are narrowed down to $3.34^{+1.0}_{-0.7}$ kpc \citep{Bailer2021}, which is consistent with the SNR distance. 
Furthermore, the spectroscopic study by \citet{Ihara2007} reported that star E is the only star in our sample to exhibit an absorption Fe I line at 3720 $\AA$ \citep[though this detection is disputed by][]{Gonzalez2009}. The fact that only the blueshifted side of the absorption feature was detected would indicate that the star is within the SNR sphere. However, confirming whether star E resides inside the SNR or is located in its background is challenging due to uncertainties associated with the template stellar spectra that impact the detection of the redshifted side of the absorption feature \citep{Ihara2007}.

For star B, \citet{Kerzendorf2018} and \citet{Ruiz-Lapuente2019} ruled out an association with the SNR based on several arguments. One of them was its distance, but the \emph{Gaia} DR2 parallax used in these papers has evolved from 0.491$\pm$0.051 mas ($\sim$ 2 kpc) to 0.373$\pm$0.032 mas in DR3; thus this would place the star slightly further away ($2.53^{+0.23}_{-0.20}$ kpc) and in better agreement with the UV-optical luminosity distance estimate of $d = 2.63 ^{+0.69}_{-0.23} $ kpc from \citet{Kerzendorf2018} using \textit{Hubble} space telescope data.

As the stellar companion in a Type Ia explosion is supposed to have been flung out of the system, the remaining donor star after the supernova is expected to have an unusual velocity with respect to surrounding stars.
Therefore, we compared the velocity properties of star B (V= 54 km s$^{-1}$ at a distance of 2.53 kpc) with the sample of stars in a 30' radius, lying in a distance slice of 2.5-4 kpc, a parallax fractional error better than 20 $\%$ (good distance estimate), and a proper motion error better than 0.1 mas yr $^{-1}$.  This sample resulted in a total of $\sim$500 stars. When building a histogram of stellar tangential velocities, estimating the velocity for each star at its geometrical distance, star B is within the 25th percentile of fastest stars in this sample.
In theory this exercise should be carried out using the full 3D stellar velocity of the sample.
 However, while the radial velocity of star B has been measured  \citep[V$_{\rm rad} = 51.29 \pm 1.8$ km s$^{-1}$, ][]{Kerzendorf2018}, only $\sim$150 out of our 500 stars have  Gaia radial velocity measurements.
In this biased sample (mostly limited by magnitude), the 3D velocity of star B (V=74 km s$^{-1}$) is below the median value (85 km s$^{-1}$) of the sample, showing that it has no particular velocity with respect to the neighboring stars.

In light of the latest measurements from \emph{Gaia} DR3, it appears that stars B and E are the only potential donor stars for the SNR, with the other stars likely being foreground objects. Thus, it may be concluded that either star B is associated with SN 1572 in a single degenerate scenario, wherein most of the Fe inside is highly ionized to account for the absence of an Fe II absorption line in its UV spectrum \citep{Kerzendorf2018}. Alternatively, star E could be the progenitor, but further spectroscopic observations are required to confirm the Fe I absorption feature. Finally, it is possible that there is no discernible stellar progenitor and that SN 1572 resulted from a double degenerate explosion.

\begingroup
\renewcommand{\arraystretch}{1.4} 
\begin{table}[h!]
\caption{\footnotesize Properties of SN 1572 potential donor stars from the \emph{Gaia} DR3 catalog \citep{Gaia22}. The geometrical distance derived from the parallax is given at the 16th, 50th, and 84th percentiles from the posterior distribution of the Baeysian distance estimate from \citet{Bailer2021}. Mag is the G-band mean magnitude.}             
\centering                          
\begin{tabular}{c|c c c c c c }        
\hline\hline                 
Star & Mag & Parallax (mas) & d$_{\rm geom}$ (kpc)   \\    
\hline                        
A  & 12.41 & 0.825 $\pm$ 0.035 &  1.20$^{+0.06}_{-0.04}$  \\ 
B  & 15.11 & 0.373 $\pm$ 0.032 &  2.53$^{+0.23}_{-0.20}$ \\ 
C  & 18.17 & 3.561 $\pm$ 0.523 &  0.30$^{+0.05}_{-0.04}$\\ 
D  &  19.36 & 1.256 $\pm$ 0.282 & 0.93$^{+0.27}_{-0.19}$ \\ 
E  &  18.93 & 0.266 $\pm$ 0.175 & 3.52$^{+2.0}_{-1.0}$ \\ 
G  &  17.96 & 0.518 $\pm$ 0.099  & 1.95$^{+0.47}_{-0.32}$  \\ 
\hline                                   
\end{tabular}
\label{tab:gaia}
\end{table}
\endgroup

\subsection{3D reconstruction}
\label{subsection:3D}

\begin{figure*}
\centering
\includegraphics[scale=0.8, trim = 43 375 43 40, clip=true]{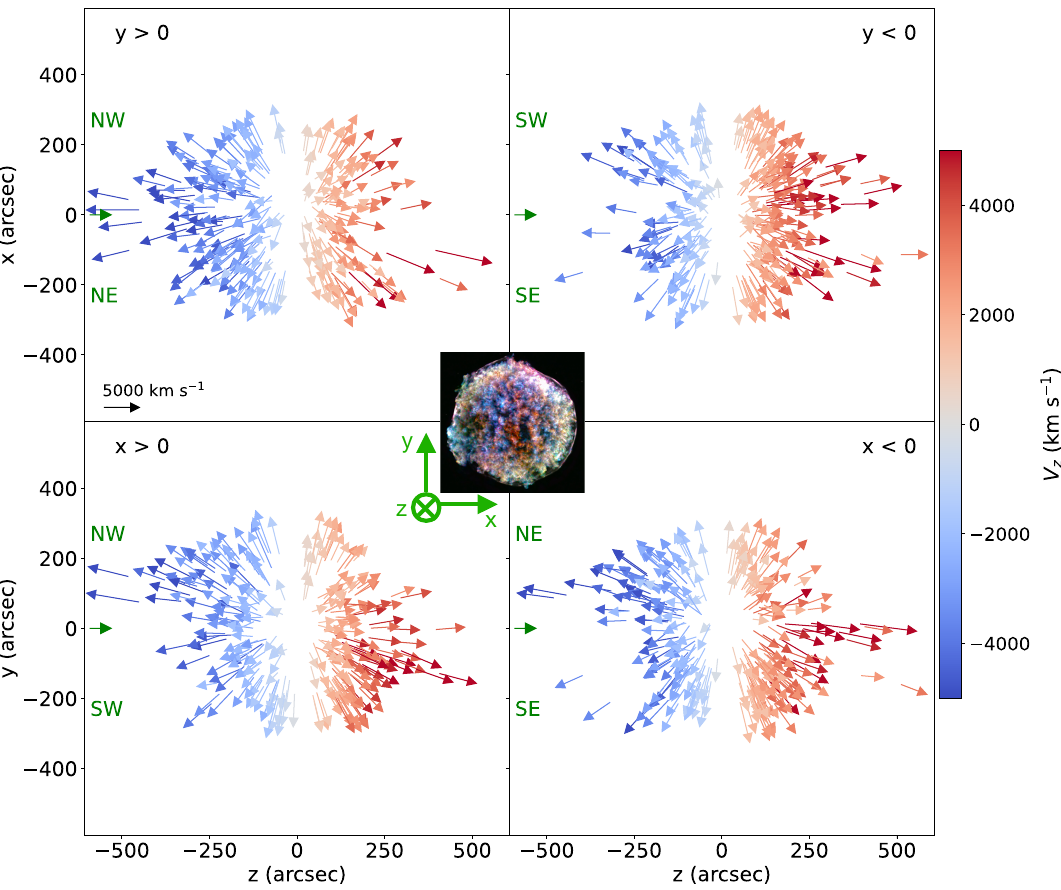}
\caption{\footnotesize 3D vector field of the dynamics of Tycho's SNR based on our results. \emph{Top right and left:} View along the y-axis (from above). \emph{Bottom right and left:} View along the x-axis (from the right). The colors are the velocities $V_{\rm z}$. The green arrows at left indicate the position of the observer and we added (in green) some indications of the zones seen by the observer looking at this plot. The lack of vectors for positions in the LoS $z$ near zero, is due to a selection bias (see the text). A 3D visualisation is available online.}
\label{vues3D}
\end{figure*}

In this study, we obtained the velocities in the plane of the sky, $V_{\rm x}$ and $V_{\rm y}$, and the integrated velocity in the LoS, $V_{\rm z}$. We have also directly pinpointed the position of the vector in the plane of the sky ($x$ and $y$). Two limitations remain to obtain a complete 3D reconstruction of the SNR and its dynamics. We need the $V_{\rm z}$  in one point (not an integration over the LoS) and the position of this point in the LoS, $z$.

To limit this integration on the LoS problem, we selected only the regions that are dominantly red or blue-shifted. To do this we detect local extrema on our map of $V_{\rm z}$ velocity  using the tool {\tt peak\_local\_max} \footnote{\url{https://scikit-image.org/docs/stable/auto_examples/segmentation/plot_peak_local_max.html}} of the \emph{Skimage} library. Choosing points that are not too near to the SNR edge, where $V_{\rm z}$ is poorly determined, we obtained around 350 redshifted points and 320 blueshifted points, evenly spaced on the SNR. Then we applied the tool POF presented in Section \ref{subsection:Optical Flow to measure proper motion} to measure the corresponding proper motion of these specific features.
Finally, we selected only the points with an expansion index $m$ less than 1.2 (as in Section \ref{subsection : The vector field}) and an angular deviation from a radial expansion less than 40,° ending up with a collection of nearly 530 points.
For this sample, the mean of the space velocity, $V_{\rm xyz}$, is 3650 km s$^{-1}$, with a standard deviation of 1420 km s$^{-1}$. This is in agreement with the values found by \cite{Millard2022}, namely, in the range of around 1900-6000 km s$^{-1}$.

To obtain the  LoS position, $z$, for each of these points, we must add a hypothesis. If we suppose that the velocity and radius vectors are colinear, there is a simple kinematic relation \( z = \frac{V_{\rm z}}{V_{\rm xy}} \ r_{\rm xy}\) that is true in each point. In Section \ref{subsection : The vector field}, we obtain an estimation of the angular deviation between the radius and the velocity vectors: its distribution is a Gaussian centered around zero with a standard-deviation of around 20°. So, the position $z$ that we obtain is only an approximation.
Nevertheless, we have now a proxy of the space radius, $r_{\rm xyz}$, for all of our points. Overall, 75\% of our sample has a space radius bigger than 2.2 arcmin. That value is between the estimation of the position of the reverse shock from \cite{Yamaguchi2014} of 2.6 arcmin and the one from \cite{Millard2022} of 2.0 arcmin.

Finally, combining the parameters ($x$, $y$, $z$) and the 3D velocity vectors,  we obtained a full reconstruction of the dynamics of Tycho's SNR, presented in Fig. \ref{vues3D}. Each plot represents the expansion of half a shell viewed along the x or y-axis. Each arrow is color coded with its LoS $V_{\rm z}$. The lack of vectors for $z$ around 0 is due to the local extrema search, which selects only high V$_{\rm{z}}$ values.
In Section \ref{subsection:ratiotovelocity}, we also underline that due to calibration uncertainties, the values of  V$_{\rm{z}}$ with a norm less than 500 km s$^{-1}$ are unreliable. Broadly speaking, we must be aware that the distribution of our sample is not evenly distributed in the emission shell, so we must be cautious with the zones where there is a lack of vectors.

The north-south asymmetry we see in the integrated $V_{\rm z}$ map (Fig. \ref{CarteEc}) is visible in these 3D views: in the top left, the north half  of the shell is more blueshifted and on the contrary, the south half shell is dominantly redshifted (top right panel). In the same way, in the bottom panels, a bipolar large-scale velocity asymmetry is observed, which agrees with this large-scale asymmetry. 

To obtain an interactive representation of this complex dataset, we used the tool \emph{Blender}\footnote{\url{https://www.blender.org}} to build a 3D visualization of our results, which can be found in the platform \emph{Sketchfab} \footnote{ \href{https://skfb.ly/oKYZp}{Online visualisation of our 3D vector field} }


\section{Conclusion}

The \emph{Chandra} observations of the Tycho supernova remnant are the perfect dataset for the application of new tools to study the SNR ejecta dynamics in greater detail. In the present study, we measure separately the velocity in the LoS (V$_{\rm{z}}$) and the proper motions (V$_{\rm{xy}}$) in the plane of the sky.

To estimate V$_{\rm{z}}$, we used the  GMCA tool to decompose our data cube ($x$, $y$, $E$) and to separate the redshifted and blueshifted emission. We obtained a map of the mean velocity V$_{\rm{z}}$ with full coverage of the SNR at 2" spatial resolution for the first time. 
Then, we developed the  POF tool to measure the shift of features between epochs with a 2D fit adapted to  the Poisson noise. The result for Tycho's SNR is a velocity vector field with more than 1700 vectors.
These velocity fields with an unprecedented level of detail underline the complex dynamics of Tycho's SNR despite its overall regular shape.
Our main findings in this study are  as follows:

\begin{itemize}
    \item  In the LoS, the full coverage of the SNR confirms the north-south velocity asymmetry hinted by \cite{Millard2022}. This bipolar structure could be due to an asymmetric elongated  explosion tilted towards the observer or to an interaction with some overdensity in front and behind the remnant.
    \item In the plane of the sky, a slow down of the forward-shock velocity was previously measured in the east compared to the west and associated to a gradient of density. In the ejecta dynamics, we observe that the velocity linearly increases with the radius in the western undisturbed region, with a forward shock that is faster than the ejecta. Meanwhile, in the east, the dynamics are more complex, likely due to the density gradient, and some inner ejecta have higher velocities than the forward shock.
    
    \item At small scales, we observe in our $V_{\rm xy}$ vector field an interesting structure in the north-east, where the velocity increases followed by a decrease with an increasing radius. 
    This is unexpected as the velocity profile should increase linearly with radius due to projection effects.
    The position of this feature matches a potential molecular cloud seen in the radio. This could be interpreted as a complex projected profile of the current deceleration of the ejecta interacting with the cloud or two different components, such as a fast knot in the foreground and ejecta slowing down in the background.
    \item Using the V$_{\rm{xy}}$ field of ejecta vectors, we estimated the center of explosion by finding the common origin of these vectors and we revisited the properties of potential stellar progenitors using the \emph{Gaia} DR3 catalog. The latest parallax measurements place stellar candidate B slightly further away (d$\sim$2.5 kpc) than in the DR2 catalog (d$\sim$2.0 kpc), at a distance now compatible with the SNR. With improved measurements,  star E is also an interesting alternative candidate at a distance of $\sim$3.5 kpc and with potential Fe I line absorption due to the SNR ejecta.
    
    \item Combining V$_{\rm{xy}}$ and  V$_{\rm{z}}$, we reconstructed a 3D vector field with around 450 positions in the SNR and built an interactive vizualization of this complex dataset. 
\end{itemize}

The new methods developed in this study benefit from the very good statistics of the data of Tycho's SNR observed by the \emph{Chandra} telescope. However, they could also be used for other supernova remnants or astrophysical objects. In particular, the GMCA algorithm is a powerful tool for decomposing any cube of data with good contrast between the underlying components. The tool POF could be also applied on other objects to study their dynamics if they are observed with sufficient spatial resolution and statistics.

\begin{acknowledgements}
We thank Gabriel Pratt for his useful comments on the analysis and on the manuscript, and Jérôme Bobin for discussions on the GMCA method. We also thank Benjamin Romain, who produced the 3D visualisation with \emph{Blender}.
The research leading to these results has received funding from the European Union’s Horizon 2020 Programme under the AHEAD2020 project (grant agreement n. 871158). This work was supported by CNES, focused on methodology for X-ray analysis.
This work has made use of data from the European Space Agency (ESA) mission
{\it Gaia} (\url{https://www.cosmos.esa.int/gaia}), processed by the {\it Gaia}
Data Processing and Analysis Consortium (DPAC,
\url{https://www.cosmos.esa.int/web/gaia/dpac/consortium}). Funding for the DPAC
has been provided by national institutions, in particular the institutions
participating in the {\it Gaia} Multilateral Agreement.
\end{acknowledgements}


\begin{appendix} 

\section{Poisson optical flow}
\label{Appendix : POF}

\begin{figure*}[h]

\begin{subfigure}{1\textwidth}
\centering
\includegraphics[scale=0.4, trim = 0 0 0 0,clip=true]{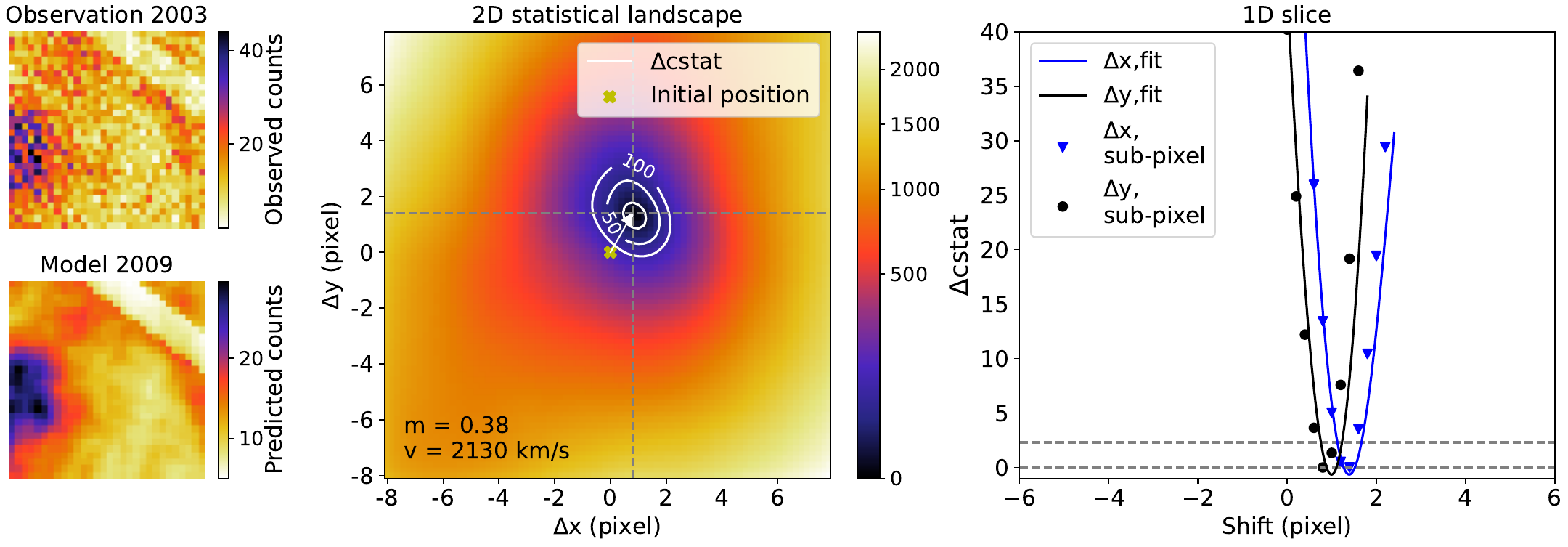}
\label{fig:subim4}
\end{subfigure}

\begin{subfigure}{1\textwidth}
\centering
\includegraphics[scale=0.4, trim = 0 0 0 0,clip=true]{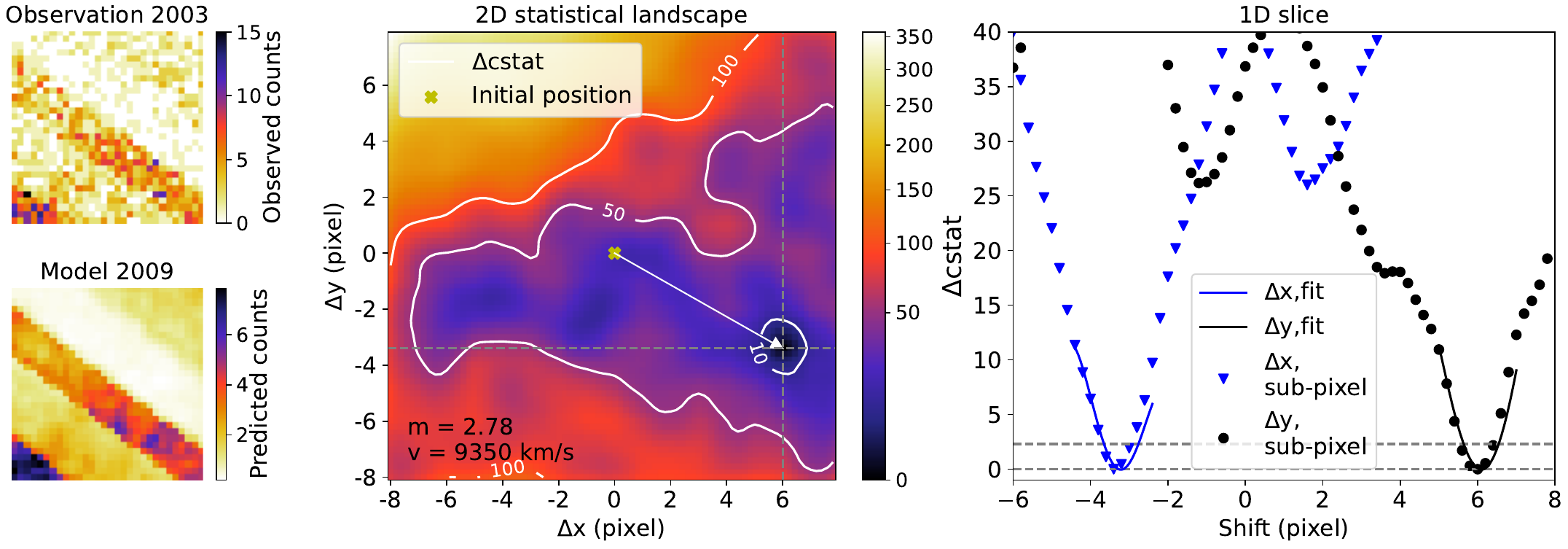} 
\label{fig:subim1}
\end{subfigure}

\begin{subfigure}{1\textwidth}
\centering
\includegraphics[scale=0.4, trim = 0 0 0 0,clip=true]{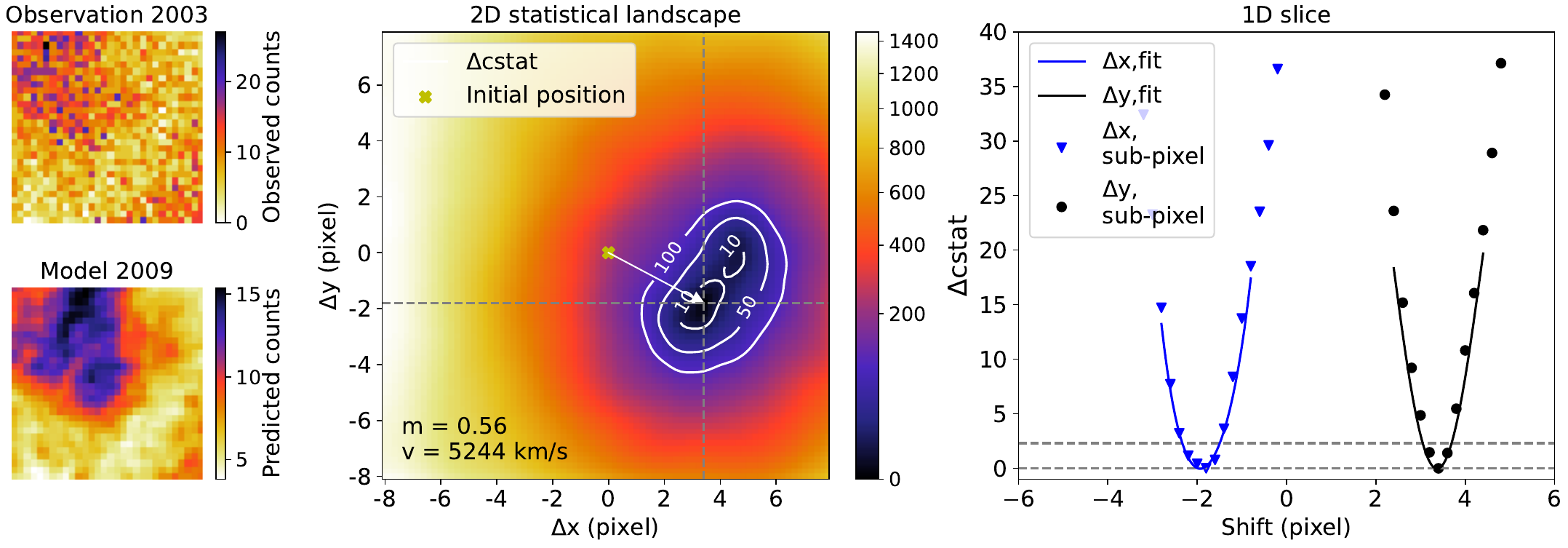}
\label{fig:subim2}
\end{subfigure}

\begin{subfigure}{1\textwidth}
\centering
\includegraphics[scale=0.4, trim = 0 0 0 0,clip=true]{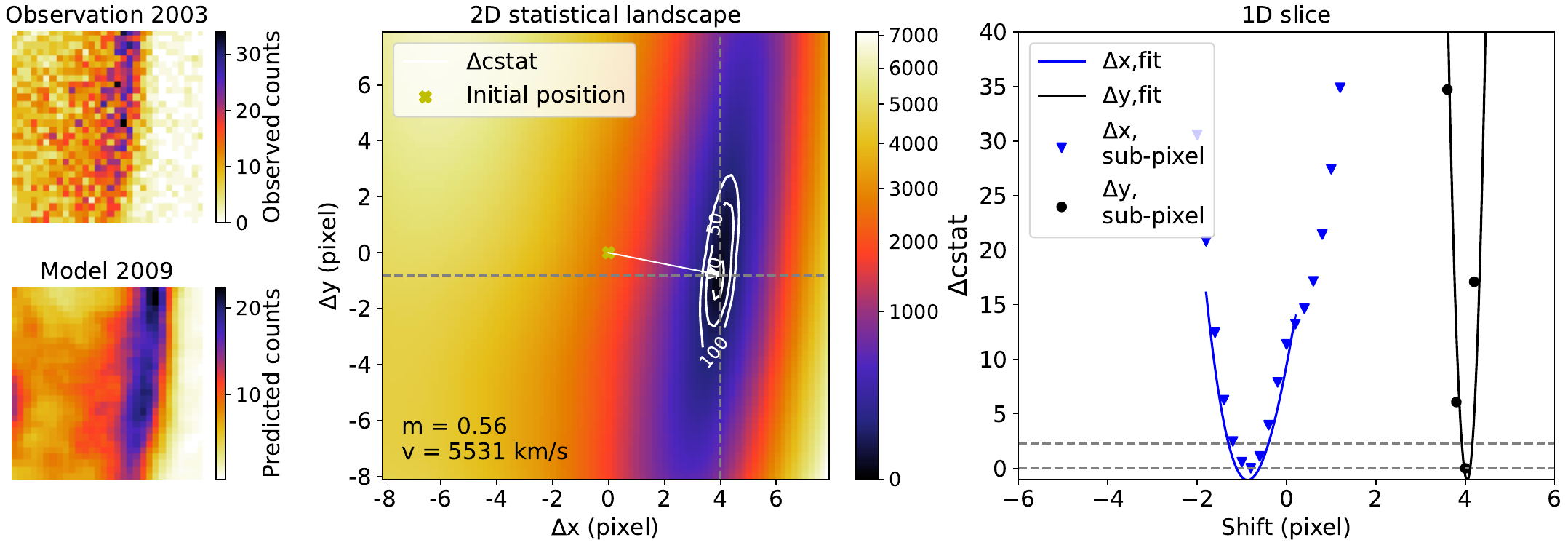}
\label{fig:subim3}
\end{subfigure}

\caption{\footnotesize Four detailed examples of our POF tool. \emph{Top left panel:} Observation vignette (data of 2003). \emph{Bottom left panel: }Model vignette (data of 2009) smoothed by $0\farcs4$, corrected by the exposure map of the observation image (see Eq. \ref{EqcstatLandscape}) and without any shift. \emph{Central panel:} Statistical landscape obtained when we shift the model vignette of ($\Delta$x, $\Delta$y). The values have been subtracted from the the minimum value of the landscape. The output vector, velocity, and expansion index are indicated. \emph{Right panel:}  Slices of this landscape (blue triangles along the x-axis and black dots along the y-axis) and the slices of the interpolating 2D polynomial. The horizontal lines correspond to the minimum of the landscape and the $\Delta$cstat value to have 1 sigma uncertainties.}
\label{fig:profilPOF}
\end{figure*}

The principle of our  POF tool is presented in Section \ref{subsection:Optical Flow to measure proper motion}, we add in this Appendix some details about its strengths and limitations. Figure \ref{fig:profilPOF} summarizes the steps of the algorithm: the initial vignettes from the observation (data from 2003) and the model (data from 2009), the statistical landscape obtained from Eq. \ref{EqcstatLandscape} and some slices of this landscape.

The first example is a good case, with contrasted features. Nevertheless, it confirms the necessity of the statistical landscape's interpolation to obtain the full uncertainty ellipse. Even with a subpixel step during the construction of the statistical landscape (blue and black markers in the right panels of Fig. \ref{fig:profilPOF}), the $\Delta$cstat corresponding to a 1 sigma uncertainty lies between these points. We can note that the continuous lines in the right panels are slices of the 2D fit of the statistical landscape and so, this is why they do not perfectly match the points. 

This example illustrates also the correction of the exposure map. As mentioned, the observation vignette is not modified at any step to keep its Poisson statistic. The model vignette is shifted and then the exposure map of the observation vignette is applied to the model and morphological information in the CCD gaps is mostly lost. For the first example this loss of information is not dramatic but the second sub-figure stresses the worst case of statistical landscape. Therefore a selection of the outputs (the expansion index, $m,$ must be inferior to 1.2) is necessary to avoid these poorly constrained vectors.

Finally, the last examples exhibit some limitations of our method. Depending on the morphology of the feature and its variability between epochs, some local minima can compete in the statistical landscape. Some morphologies are also not adapted to measure a 2D shift without additional constraints. For the synchrotron filaments, there is a degeneracy along the filaments, thus, our tool can struggle to follow the forward shock if it is purely linear in shape.

\section{Energy at rest of the Si line}
\label{Appendix : Energy at rest}

\begin{figure*}
\centering
\includegraphics[scale=0.4, trim = 0 0 0 0,clip=true]{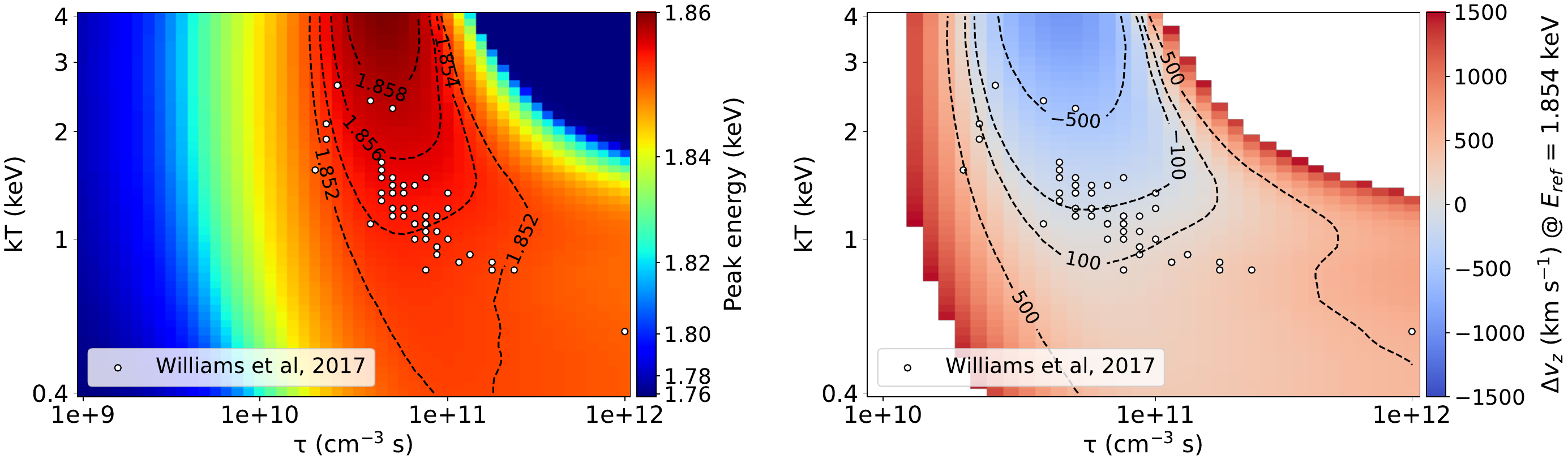}
\caption{\footnotesize Impact of the local conditions of temperature and ionization timescale on the energy at rest of the silicon line used to obtain the velocity in the LoS, $V_{\rm z}$. \emph{Left: }  Background map is the theoretical peak energy of the silicon line in the (k$T$, $\tau$) space with contours in black. This is compared with the k$T$ and $\tau$ local spectral measurements from \cite{Williams2017}. \emph{Right:} Assuming a LoS velocity of 4000 km s$^{-1}$, these panels show the associated velocity uncertainty if the energy of reference changes due temperature and ionization timescale variations.}
\label{carte_kT_tau}
\end{figure*}

To transform the peak energy map of the silicon line in a map of velocity along the LoS, we need a reference, namely, the  energy of the line at rest. This value depends on the local (x, y, z) conditions of temperature and ionisation. These conditions vary in the shell and in the remnant according to the local interaction with the circumstellar medium, the speed of the forward shock and/or reverse shock, and the time since being shocked. In our case, we must suppose that the energy of reference is the same in all the remnant because there is no map of the temperature and ionization timescale available to create a reference map at the pixel level. We take the value $E_{ref} = 1.854$ keV as explained below.

To evaluate the value of our reference and the impact of the variation of temperature (k$T$ in keV) and ionisation timescale ($\tau$ in $\rm{cm^{-3}} s$), we created a library of spectra with the model \emph{nei} (non equilibrium ionisation) of \emph{Xspec} using AtomDB v3.0.9. We formed a grid of spectra with fifty temperatures between 0.4 keV and 4 keV and fifty ionisation timescales between $10^{9}$ and $10^{12}$ $\rm{cm^{-3}} s$ logarithmically distributed. These spectra are based on the response files \emph{ARF} and an \emph{RMF} from the observation of Tycho's SNR in 2009 by the \emph{Chandra} telescope.
Then for each of these spectra, the precise maximum of the line in the band 1.75-1.95 keV was obtained thanks to a local polynomial interpolation. We stack these results in the Fig. \ref{carte_kT_tau} (left), where the color codes for the value of the peak energy. In this figure, we add the measures of temperature and ionisation timescale obtained by \cite{Williams2017} with their 57 zones where they fit the redshift and these parameters. Their measures are representative of the entire remnant and lie in a "valley" in our landscape (k$T$, $\tau$) with a stable value for the peak energy between 1.852 and 1.856 keV. So we choose an energy of reference of 1.854 keV.

Then we transform this map (k$T$, $\tau$, $E_{\rm{ref}}$) in a velocity map in order to know the impact of the variation of physical conditions at rest if we suppose a constant reference in all the SNR. This new map is shown at right in the Fig. \ref{carte_kT_tau}, and covers only the zone of interest to have a more readable colorbar. We can see that the variations of velocity due to temperature and ionisation time are in general less than 500 km s$^{-1}$, and a bit more for high temperature or high ionisation time.
 This is marginal in comparison with our $V_{\rm z}$ velocities, which have an order of magnitude of $10^3$ km s$^{-1}$. To take into account this bias, we added an uncertainty of 500 km s$^{-1}$ to all our velocities on the LoS.

\end{appendix}

\end{document}